\documentstyle[mnras_cite,psfig]{mn}

\def\gsim{~\rlap{$>$}{\lower 1.0ex\hbox{$\sim$}}}
\def\lsim{~\rlap{$<$}{\lower 1.0ex\hbox{$\sim$}}} 
\def\d{{\rm d}}

\begin{document}

\title[Self-Consistent Theory of Halo Mergers]{Self-Consistent
Theory of Halo Mergers}
\author[Andrew~J.~Benson, Marc
Kamionkowski \& Steven H.~ Hassani]{Andrew~J.~Benson$^1$, Marc
Kamionkowski$^2$ \& Steven H.~Hassani$^{2,3}$ \\
$^1$Department of Physics, University of Oxford, Keble Road,
Oxford OX1 3RH, United Kingdom (e-mail: {\tt abenson@astro.ox.ac.uk})
\\ $^2$Mail Code 130-33, California Institute of Technology,
Pasadena, CA~91125, U.S.A. (e-mail: {\tt
kamion@tapir.caltech.edu})\\
 $^3$Department of Physics, Princeton University, Princeton,
 NJ~08544, U.S.A. (email: {\tt shassani@princeton.edu})}

\maketitle

\begin{abstract}
The rate of merging of dark-matter halos is an absolutely essential
ingredient for studies of both structure and galaxy formation.
Remarkably, however, our quantitative understanding of the halo merger
rate is still quite limited, and current analytic descriptions based
upon the extended Press-Schechter formalism are fundamentally
flawed. We show that a mathematically self-consistent merger rate must
be consistent with the evolution of the halo abundance in the
following sense: The merger rate must, when inserted into the
Smoluchowski coagulation equation, yield the correct evolution of the
halo abundance. We then describe a numerical technique to find
merger rates that are consistent with this evolution. We present
results from a preliminary study in which we find merger rates that
reproduce the evolution of the halo abundance according to
Press-Schechter for power-law power spectra. We discuss the
limitations of the current approach and outline the questions that
must still be answered before we have a fully consistent and correct
theory of halo merger rates.
\end{abstract}

\section{Introduction}

In current cosmological theory the mass density of the Universe is
dominated by dark matter. The most successful model
of structure formation is that based upon the concept of cold dark
matter (CDM). In the CDM hypothesis dark-matter particles interact
only via the gravitational force. Since the initial distribution of
density perturbations in these models has greatest power on small
scales, the first objects to collapse and form dark-matter halos are of
low mass. Larger objects form through the merging of these smaller
sub-units. Consequently, the entire process of galaxy formation is
thought to proceed in a ``bottom-up'', hierarchical manner.

Clearly then, the rate of dark-matter--halo mergers is an absolutely
crucial ingredient in models of galaxy and large-scale-structure
formation, from sub-galactic scales to galactic and galaxy-cluster
scales. The Press-Schechter (PS) formalism \cite{PreSch74} has long
provided a simple, intuitive, and surprisingly accurate formula for
the distribution of halo masses at a given redshift over a large range
of mass scales and for a vast variety of initial power spectra. This
formalism states that the number of halos per comoving volume with
masses in the range $M\rightarrow M+\d M$ is \cite{PreSch74}
\begin{eqnarray}
    n(M;t) \d M &=& \sqrt{2\over \pi} {\rho_0 \over M^2} {\delta_c(t)
     \over \sigma(M)} \left|{ \d \ln \sigma \over \d\ln M}
     \right| \nonumber \\
     & & \times \exp\left[ - {\delta_c^2(t) \over 2 \sigma^2(M)}\right] \d M,
\label{eqn:PSabundance}
\end{eqnarray}
where $\rho_0$ is the background density and $\delta_{\rm c}(t)$ is
the critical overdensity for collapse in the spherical-collapse
model. Here, $\sigma(M)$ is the root variance of the primordial
density field in spheres containing mass $M$ on average, extrapolated
to $z=0$ using linear theory; it can be determined from the primordial
power spectrum $P(k)$ that is specified by inflation-inspired models
for primordial perturbations.  For power-law power spectra, $P(k)
\propto k^n$ and $\sigma(M) \propto M^{-(3+n)/6}$; in this case, the
Press-Schechter halo abundance diverges as $n(M)\propto M^{(-9+n)/6}$
as $M\rightarrow0$, and it is exponentially suppressed above a
characteristic mass $M_*(t)$ determined from the condition
$\delta_c(t)=\sigma(M_*)$.  The critical overdensity $\delta_c(t)$ is
a monotonically decreasing function of $t$ so that the Press-Schechter
distribution shifts to larger masses with time, i.e. $\dot{M}_*>0$.
The fraction of cosmological mass in halos of mass $M\rightarrow M+\d
M$ is $Mn(M)\d M/\rho_0$, and, at any given time, most of the mass
resides in halos with masses $M\sim M_*$.  Note that although the
number of halos diverges as $M\rightarrow0$, the total mass in halos
remains finite.

An elegant paper by \scite{LacCol93}---and similar work by
\scite{Bonetal91} and \scite{Bow91}---extended the work of Press and
Schechter to determine the rate at which halos of a given mass merge
with halos of some other mass. In addition to providing valuable
physical insight, these merger rates have extraordinary practical
value, having been applied to galaxy-formation models, e.g., if galaxy
morphologies are determined by the merger history \cite{GotKlyKra99};
AGN activity \cite{WyiLoe03}; models for Lyman-break galaxies
\cite{Koletal99}; abundances of binary supermassive black holes
(SMBHs) \cite{VolHaaMad02}; rates for SMBH coalescence \cite{MilMer01}
and the resulting LISA event rate \cite{MenHaiNar01,Hae94}; the first
stars \cite{SanBroKam02,ScaSchFer03}; galactic-halo substructure
\cite{KamLid00,BulKraWei00,Benetal02,Som02,StiWidFri01}; halo angular
momenta \cite{Vivetal02} and concentrations \cite{Wecetal02}; galaxy
clustering \cite{Peretal03}; particle acceleration in clusters
\cite{GabBla03}; and formation-redshift distributions for galaxies and
clusters and thus their distributions in size, temperature,
luminosity, mass, and velocity \cite{Veretal01,VerHaiSpe02}.

Amazingly enough, however, these merger rates are fundamentally
flawed.  As we show below, the extended-Press-Schechter (EPS) formulae
for merger rates are mathematically self-inconsistent, providing {\it
two} different results for the same merger rate.\footnote{We first
discovered this inconsistency in \scite{SanBroKam02}.}  The two
different merger rates are plotted in Fig.~\ref{fig:PSprob}. They are
equal for equal-mass mergers but increasingly discrepant for larger
mass ratios.  This ambiguity will be particularly important for, e.g.,
understanding galactic substructure and for SMBH-merger rates. Even
the smaller numerical inconsistency for mergers of nearly equal mass
may be exponentially enhanced during repeated application of the
formula while constructing merger trees to high redshift. Moreover,
the ambiguity calls into question the entire formalism, even when the
two possibilities seem to give similar answers
quantitatively\footnote{Extended Press-Schechter theory discusses the
correlation of peaks in the primordial mass distribution.  It is the
association of such peaks with bound halos, which is not necessarily
well-defined, that leads to these problems with the derived merger
rates.}.

In this paper, we discuss the mathematical requirements of a
self-consistent theory of halo mergers.  As recognised already
\cite{SilWhi78,She95,ShePit97}, the merger process is described by the
Smoluchowski coagulation equation. This equation simply says that the
rate at which the abundance of halos of a given mass changes is
determined by the difference between the rate for creation of such
halos by mergers of lower-mass halos and the rate for destruction of
such halos by mergers with other halos. The correct expression for the
merger rate must be one that yields the correct rate of evolution of
the halo abundance when inserted into the coagulation equation The
problem is thus to find a merger rate, or ``kernel,'' that is
consistent with the evolution of the halo abundance, either the
Press-Schechter abundance or one of its more recent N-body--inspired
variants \cite{SheTor99,Jenetal01}.

The apparent simplicity of the mathematical problem, which appears in
equation (\ref{eq:Smoluchowski}) below, is in fact quite
deceptive. The Smoluchowski coagulation equation is in fact an
infinite set of coupled nonlinear differential equations. The equation
appears in a variety of areas of science---e.g., aerosol physics,
phase separation in liquid mixtures, polymerization, star-formation
theory \cite{AllBas95,SilTak79}, planetesimals
\cite{Wet90,MalGoo01,Lee00}, chemical engineering, biology, and
population genetics---so there is a vast but untidy literature on the
subject (although see \pcite{leyvraz} for an illuminating review). It
has been studied a little by pure and applied mathematicians
\cite{Ald99}. Still, solutions to the coagulation equation are poorly
understood.  Furthermore, there is virtually no literature on the
problem we face: i.e., how to find a merger kernel that, when inserted
into the coagulation equation, yields the desired halo mass
distribution and its evolution as a
solution.

In this paper, we present a numerical technique to find a merger
kernel that yields the correct evolution of a specified halo mass
distribution. We demonstrate this technique by applying it to
Press-Schechter distributions for power-law power spectra. We regret
that at this point we still do not have results that can be applied to
astrophysical merger rates (e.g. those valid for CDM power spectra),
although our techniques can easily be extended to more realistic
cases. Moreover, although we have indeed found merger rates that are
mathematically consistent with the desired halo distributions, our
inversion of the Smoluchowski equation is not necessarily unique. As
we discuss below, there may be other merger kernels that also yield
the same halo mass distribution. More work must be done to determine
how to insure that the numerical inversion yields the merger kernel
that in fact describes the process of mergers from gravitational
clustering of mass with an initial Gaussian distribution.
Nevertheless, the work presented here may be a first step in this
direction.

Below, in \S\ref{sec:review}, we first review the extended
Press-Schechter calculation and show that it gives mathematically
inconsistent expressions for the merger rate.  In Section 3, we then
discuss the coagulation equation that needs to be solved.  Section 4
describes our numerical algorithm for finding self-consistent merger
rates.  Section 5 and Figs. 2--8 show results of our numerical
inversion for a variety of power-law power spectra.  Section 6 answers
some common questions about this work, and Section 7 provides some
concluding remarks and outlines some questions that must still be
addressed in future work.  Appendix A provides an alternative
formulation of the coagulation equation that makes the cancellation of
divergences explicit.  Appendix B provides, for reference, derivations
of the two fo the known analytic solutions to the coagulation
equation.

\section{Review of the Extended Press-Schechter Calculation}
\label{sec:review}

\begin{figure}
\psfig{file=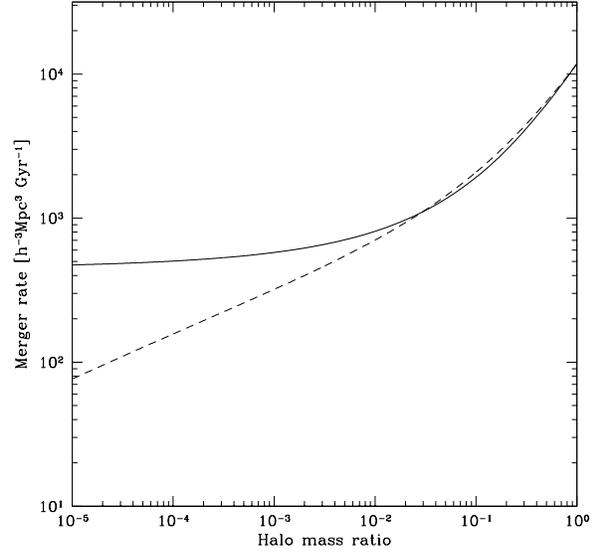,width=80mm}
\caption{The dark-matter-halo merger rate computed using the extended
Press-Schechter formula of \protect\scite{LacCol93}. Results are shown
for an $n=1$ power-law power spectrum, a halo of mass
$M_1=10^{12}h^{-1}M_\odot$ at $z=0$ in a Universe with $\Omega_0=0.3$,
$\Lambda_0=0.7$, $\sigma_8=0.9$ and $\Gamma=0.21$. The two lines show
the result of using the two versions of the merger rate predicted by
the extended Press-Schechter theory.}
\label{fig:PSprob}
\end{figure}

The extended Press-Schechter theory
\cite{LacCol93,LacCol94,Bonetal91,Bow91} predicts the distribution of
masses of progenitor halos for a halo of a given mass. By manipulating
the equations of this theory it is possible to obtain an expression
for the merger rate of halos of mass $M_1$ with those of mass
$M_2$. \scite{LacCol93} give the following expression for the
probability per unit time that a halo of mass $M_2$ will merge with a
halo of mass $M_1=M_{\rm f}-M_2$ to make a halo of mass $M_{\rm f}$
\begin{eqnarray}
  {\d^2 p \over \d M_2 \d t} &=& {1\over M_{\rm f}} \sqrt{2 \over
     \pi} \left| {\dot \delta_c \over \delta_c} \right| \left| { \d\ln \sigma(M_f) \over
     \d\ln M_f} \right| {\delta_c(t) \over \sigma(M_f)} \nonumber
     \\
     & \times & {1\over \left[1-\sigma^2(M_f)/\sigma^2(M_1)\right]^{3/2}} \nonumber \\
 & \times &
     \exp\left [ -{1\over2} \delta_c(t) \left( {1\over
     \sigma^2(M_f)} - {1\over \sigma^2(M_1)} \right) \right].
\label{eq:prob}
\end{eqnarray}
In the extended-Press-Schechter formalism, the abundance of halos of
mass $M$ is still given by equation (\ref{eqn:PSabundance}). The total
number of mergers between halos of mass $M_2$ and $M_1$ per unit time
and per unit volume must therefore be
\begin{equation}
     R(M_1,M_2;t) = n(M_2;t) {\d^2 p \over \d M_2 \d t}.
\label{eq:definerate}
\end{equation}
Since the merger rate is proportional to both $n(M_1;t)$ and
$n(M_2;t)$ we define a merger kernel $Q(M_1,M_2;t)$ (which has
units of cross section times velocity) through
\begin{equation}
     R(M_1,M_2;t) = Q(M_1,M_2;t) n(M_1;t) n(M_2;t).
\label{eq:rate}
\end{equation}
From equations~(\ref{eqn:PSabundance})--(\ref{eq:rate}) we can derive
\begin{eqnarray}
     Q(M_1,M_2;t) &=& {M_2^2 \over \rho_0 \sigma_{\rm f}} {
     \sigma_2 \over M_2} \left|{\dot \delta_c \over \delta_c}
     \right| \left| {\d \ln \sigma_f \over \d \ln M_{\rm f}} \right|
     \left| {\d \ln \sigma_2 \over \d \ln M_2} \right|^{-1} \nonumber \\
     & \times & {1
     \over (1-\sigma_{\rm f}^2/\sigma_1^2)^{3/2}} \nonumber \\
     &\times& \exp \left[ - {\delta_c^2 \over 2} \left( {1\over
     \sigma_{\rm f}^2} - {1\over \sigma_2^2} - {1\over \sigma_1^2} \right)
     \right],
\label{eq:qeps}
\end{eqnarray}
where we have adopted the notation $\sigma_2 = \sigma(M_2)$, etc. The
problem with this merger rate is immediately apparent. Clearly
$R(M_1,M_2;t)=R(M_2,M_1;t)$ must always hold; i.e., the merger rate
must be a symmetric function of its arguments. However, this is not
the case for the above definition of $Q(M_1,M_2;t)$. This can be seen
clearly in Fig.~\ref{fig:PSprob} where we plot the merger kernels
$Q(M_1,M_2;t)$ and $Q(M_2,M_1;t)$ for a specific case. Although the
differences between the two predicted merger rates are small over most
of the range of $M_1/M_2$ plotted, it is important to note that the
discrepancies may have significant consequences for cosmological
studies. For example, the merger rate for a mass ratio of $10^{-5}$ is
uncertain by a factor of $10$. This could significantly affect the
predicted number of dwarf galaxies expected to be found within
clusters. It is very problematic for predictions of black-hole mergers
detectable with LISA, as many of the detectable signals may arise from
mergers of very unequal masses for which the numerical ambiguity is
particularly pronounced.  Furthermore, even for mergers of nearly
equal masses, where the two predictions agree more closely, repeated
application of the merger-rate formula (as occurs in the construction
of merger trees) will lead to a growing divergence between the two
predictions.

The extended-Press-Schechter merger rate is in fact only a symmetric
function of its arguments in two special cases. The first, trivial,
case is when $M_1=M_2$. The second is for the case of a distribution
of primordial densities described by a white-noise power spectrum,
$P(k)\propto k^n$ with $n=0$. In this case the merger kernel reduces
to
\begin{equation}
Q(M_1,M_2;t) = {(M_1+M_2) \over \rho_0} \left|{ \dot \delta_c
     \over \delta_c} \right|.
\end{equation}

Our aim is to find kernels $Q(M_1,M_2;t)$ which (a) are symmetric in
their arguments and (b) yield the correct evolution of the halo
distribution $\dot{n}(M)$ when inserted into the coagulation equation.
The kernel should also (c) satisfy the known statistics of
dark-matter--halo merger rates. Below, we will illustrate a numerical
algorithm that can accomplish conditions (a) and (b); we discuss the
third condition later.

\section{Basic Equations}

During the process of hierarchical clustering, halos of mass $M$ will
be created via mergers of pairs of halos of masses $M^\prime$ and
$M-M^\prime$, and they will be destroyed via mergers with halos of any
other mass, $M^\prime$. The rate at which the abundance of halos of
mass $M$ changes is therefore
\begin{eqnarray}
     \dot n(M) = {1\over 2}\int_0^M n(M^\prime) n(M-M^\prime)
     Q(M^\prime,M-M^\prime) \d M^\prime \nonumber \\
     - n(M) \int_0^\infty
     n(M^\prime) Q(M,M^\prime) \d M^\prime,
\label{eq:Smoluchowski}
\end{eqnarray}
where the dot denotes a derivative with respect to time, the first
term on the right-hand side describes halo formation, and the second
describes halo destruction.  Note that we have suppressed the explicit
dependence of $n(M;t)$ on time and the possible dependence of
$Q(M_1,M_2;t)$ on time in equation~(\ref{eq:Smoluchowski}).  This
equation is known as the Smoluchowski coagulation equation
\cite{Smoluchowski}. It appears in a variety of areas in science in
which coagulation processes occur. One astrophysical example is the
theory of planetesimal growth, in which small objects merge to form
larger objects. In almost all prior applications, the merger kernel
$Q(M,M';t)$ is specified by (micro)physical processes (note that it
has units of cross section times velocity) and the equations are then
integrated forward from some initial mass distribution $n(M,t=0)$ to
determine the mass distribution at some later time.

In our case, however, we know the ``answer,'' the mass distribution
$n(M,t)$ (either a Press-Schechter distribution, an improved version
such as the Sheth-Tormen distribution, or some other similar
distribution determined from simulations). We need to find the merger
kernel $Q(M_1,M_2;t)$ that yields the desired evolution of this mass
distribution when inserted into the Smoluchowski equation. This, as
far as we know, is an unsolved mathematical problem. In principle,
some variant of the derivation used by \scite{LacCol93} that imposes
the symmetry constraint $Q(M_1,M_2;t)=Q(M_2,M_1;t)$ might be used to
determine this merger kernel. It is in fact easy to impose the
symmetry constraint with some {\it ansatz}, such as an arithmetic or
geometric mean of the two Lacey--Cole results. However, the merger
kernel must also be consistent, within the dictates of the coagulation
equation, with the evolution of the Press-Schechter halo distribution,
and we have not yet been able to satisfy this constraint with any
analytic approach.

In the absence of an analytic solution, we attempt to find a numerical
solution to the problem: i.e., can we numerically find a merger kernel
$Q(M_1,M_2;t)$ that when inserted into the coagulation equation yields
the evolution of the PS distribution? The answer, as we discuss below,
is yes. To illustrate the technique, we restrict our attention to the
Press-Schechter mass distribution because of the simplicity of the
analytic expressions. Moreover, we restrict our attention to power-law
power spectra, $P(k)\propto k^n$, again for simplicity. This has the
additional advantage that for the case $n=0$ we have an analytic
solution to the Smoluchowski equation. However, our technique can be
applied equally well to more accurate distributions such as the
Sheth-Tormen distribution and to other power spectra.

The Press-Schechter rate of change of halo abundance is found by
differentiating equation~(\ref{eqn:PSabundance}) and is given by
\begin{equation}
     \dot{n}(M_1,t) = - \sqrt{{2\over \pi}} {\rho_0 \over M_1^2}
     {\delta_{\rm c} \over \sigma_1} \alpha \exp\left[ - {1\over 2}
     {\delta_{\rm c}^2 \over \sigma_1^2} \right] {\dot{\delta}_{\rm
     c}\over \delta_{\rm c}} \left[{\delta_{\rm c}^2 \over
     \sigma_1^2}-1\right].
\end{equation}
Shifting to a time variable $\tau = -\ln \delta_{\rm c}(t)$ and
dimensionless mass variable $z = M/M_\star(\tau)$ [where $M_\star(\tau)$ is the
characteristic nonlinear mass scale defined through the relation
$\sigma(M_\star)=\delta_{\rm c}(\tau)$] this becomes
\begin{equation}
{{\rm d}n \over {\rm d}\tau}(z_1) = \sqrt{{2\over \pi}} {\rho_0 \over M_\star} z_1^{-2} {\delta_{\rm c} \over \sigma_1} \exp\left[ - {1\over 2} {\delta_{\rm c}^2 \over \sigma_1^2} \right] \left[{\delta_{\rm c}^2 \over \sigma_1^2}-1\right].
\end{equation}
where $\alpha = |\d \ln \sigma/\d \ln M| = (3+n)/6$. For power-law
power spectra, $\sigma(M) = \sigma(M_\star) z^{-(3+n)/6}= \delta_{\rm
c} z^{-(3+n)/6}$. Therefore,
\begin{eqnarray}
     {{\rm d}n \over {\rm d}\tau}& =& \sqrt{{2\over \pi}} {\rho_0 \over
     M_\star} z_1^{(-9+n)/6} \alpha \nonumber \\
     & &\times \exp\left[ - {1\over 2}
     z_1^{(3+n)/3} \right] \left[z_1^{(3+n)/3}-1\right].
\end{eqnarray}
The Press-Schechter abundance is simply
\begin{equation}
     n(z) =  \sqrt{{2\over \pi}} {\rho_0 \over M_\star}
     z_1^{(-9+n)/6} \alpha \exp\left[ - {1\over 2} z_1^{(3+n)/3}
     \right].
\end{equation}
In these variables, the Smoluchowski equation is
\begin{eqnarray}
{{\rm d}n \over {\rm d}\tau}(z) & = & {1\over 2}\int_0^z n(z^\prime) n(z-z^\prime) q(z^\prime,z-z^\prime){\rm d}z^\prime \nonumber \\
 & & - \int_0^\infty n(z) n(z^\prime) q(z,z^\prime) {\rm d}z^\prime,
\label{eq:rewritten}
\end{eqnarray}
where we have used $q$ to denote the merger kernel in our new system
of mass and time variables. Our goal is to find $q(z_1,z_2;\tau)$.
Clearly the solution must be proportional to $\sqrt{\pi/2}
M_\star(\tau)/\rho_0 \alpha$. We therefore choose a system of units such
that $\sqrt{\pi/2} M_\star(\tau)/\rho_0 \alpha=1$ to simplify the
calculation. This choice of units removes the explicit time dependence
from our merger rate, allowing us to find a function $q(z_1,z_2)$
valid at all times. The time dependence of the merger rate is absorbed
into a time-dependent system of units instead.

Before discussing our numerical algorithm, we first point out that, for
the mass functions we are considering, there are divergences in the
creation and destruction terms in the coagulation equation that
cancel.  As $z\rightarrow0$, the mass function $n(z) \propto
z^{-\gamma}$, where $\gamma=(9-n)/6$ is between 2 and 1 for $-3<n<3$.
Thus, if $q(z_1,z_2)$ does not vanish as one of the arguments
approaches zero (which is the case for $n=0$, and as we argue below,
should also be the case more generally), then there is a non-integrable
singularity at the lower and upper limits of the creation term in
equation~(\ref{eq:rewritten}), and one at the lower limit of the
destruction term.  These divergences cancel, however, if we impose an
infinitesimal lower mass to the limits of integration.  Physically,
halos of some given (scaled) mass $z$ are being created very rapidly
by merging of halos of infinitesimally smaller mass with halos of
infinitesimal mass, but they are also being destroyed at the same rate
by merging with infinitesimal-mass objects.  Appendix
\ref{ap:alternate} derives an alternative expression for the
coagulation equation that makes the cancellation explicit.  As we will
see below, these divergences complicate the numerical inversion, as
the matrix to be inverted will have elements of vastly differing
magnitudes.

\section{Numerical Solution}

To numerically invert the coagulation equation, we deal with a
discretized version.  We divide $z$ into $N$ intervals of size $\Delta
z$, running from $z=0$ to $z=N \Delta z$, labeled by an index $i$.
Thus, $z_i=i \Delta z$, $n_i\equiv n(z_i)$, and $q_{ij}\equiv
q(z_i,z_j)$.  We further define $y_i \equiv \d n_i/\d \tau$.  The
coagulation equation is then
\begin{equation}
     y_i = \frac{1}{2} \sum_{j=1}^{i-1} n_j n_{i-j} q_{j,i-j} - n_i
     \sum_{j=1}^N n_j q_{ij}.
\label{eq:discretizedSmol}
\end{equation}
Equation (\ref{eq:discretizedSmol}) can then be re-written in a simple
matrix form ${\bf y} = {\bf K} \cdot {\bf q}$. The vector ${\bf q}$
has $N^2$ components, corresponding to the $N\times N$ array
$q(z_i,z_j)$, while the matrix ${\bf K}$ has dimensions of $N \times
N^2$. ${\bf K}$ is the kernel matrix consisting of the $n$'s in the
above equation.  To be explicit,
\begin{equation}
     y_i = \sum_{jk} K_{ijk} q_{jk},
\end{equation}
with
\begin{equation}
     K_{ijk} = n_j n_k \left( \frac{1}{2}
     \delta_{i,j+k}-\delta_{ik}\right).
\end{equation}

In practice, we determine ${\bf K}$ by integrating the terms in the
Smoluchowski equation over each discrete interval $\Delta z$, with
$q(z_i,z_j)$ linearly interpolated across this interval. This results
in cancellation of the divergent terms and is exactly correct for the
$n=0$ case, where $q(z_i,z_j)$ is everywhere linear. This matrix equation
can in principle be solved by a suitable inversion method, or by a
least-squares minimization to find ${\bf q}$. Unfortunately, it is
simple to see that the equation is ill-determined. We have $N$ linear
equations, but $N(N+1)/2$ unknowns to determine (since $q_{ij}$ is
symmetric).  As such, there will be an infinite number of possible
solutions. We are looking therefore not just for any solution, but a
sensible one.

\subsection{Regularization conditions}

Since the above equation does not uniquely define ${\bf q}$ we have to
apply regularization conditions in order to find a physically
reasonable $q_{ij}$. Our goal is to minimize the quantity $f^2 =
|{{\bf y} - {\bf K}\cdot {\bf q}}|^2$. Since this is an ill-determined
problem, we adopt a regularization condition $R_1$ and instead seek to
minimize $|{{\bf y} - {\bf K}\cdot {\bf q}}|^2$ and $R_1^2$
simultaneously. This regularization condition should encapsulate the
desired physical properties of the solution sought. Specifically, we
will require that the solution be a smooth function of its arguments,
which seems reasonable for any physically-plausible solution, and that
$q(z_1,z_2)\geq 0$ everywhere, as a negative merger rate has no
physical meaning.

Our first regularization condition is therefore
\begin{equation}
R_1^2 = \left. \left[ \int_0^\infty \int_0^\infty [\partial^2 q/\partial z_1^2]^2 + [\partial^2 q/\partial z_2^2]^2 {\rm d} z_1 {\rm d} z_2 \right] \right/ \sigma_{R_1}^2,
\label{eq:reg}
\end{equation}
where $\sigma_{R_1}$ is an adjustable parameter. This insures that the
recovered kernels will be smoothly varying, rather than rapidly
oscillating. We will return to our second regularization condition
shortly.

The Smoluchowski equation is, of course, linear in $q(z_1,z_2)$. Since
$R_1$ is also linear in $q(z_1,z_2)$ we can use straightforward linear
algebra to solve for $q(z_1,z_2)$. Since the matrix to be inverted can
be close to singular in some instances, we explore the consequences of
optimizing the resulting solution $q(z_1,z_2)$ using a simple
minimization technique. Specifically, we aim to minimize the quantity:
\begin{equation}
     f^2 = {1\over N-1} \sum_i \left({y_i - \sum_{jk} K_{ijk }q_{jk}
     \over \sigma_i}\right)^2 + R_1^2, 
\end{equation}
where $R_1$ from equation~(\ref{eq:reg}) is replaced by a suitable
discretized expression (based upon finite differencing), and the
$1/(N-1)$ scaling ensures that the relative weight given to the two
terms in the above is independent of $N$. We choose $\sigma_i = |y_i|$
to give equal fractional weight to each element in the
summation. We find that, once the values of $N$ and $\Delta z$ have been
chosen to suit the power spectrum under consideration, the
minimization of $f^2$ leads to almost no further improvement in the
solution. As such, simple matrix inversion seems adequate to find
$q(z_1,z_2)$.

However, in some instances the matrix solution will produce solutions
for which $q_{ij}<0$ for certain $i$ and $j$. These are clearly
unphysical. As the condition $q_{ij}\geq 0$ is not linear in the
$q_{ij}$'s we instead apply this condition within our minimization
routine in such cases. The solution found from matrix inversion is
used as a starting point for the minimization. We are then able to
find smooth solutions to the Smoluchowski equation which are
everywhere positive.

Throughout, we adopt $\sigma_{R_1}=1$, which produces smooth functions
$q(z_1,z_2)$ without limiting our ability to find accurate solutions
to ${\bf y} = {\bf K} \cdot {\bf q}$.  The necessary matrix inversion
is carried out using LU decomposition. To perform the minimization of
$f^2$ we use a direction-set method \cite{Bre73}. We enforce symmetry
of the function $q(z_1,z_2)$ by re-writing the linear equations in
terms of the $N(N+1)/2$ independent components of ${\bf q}$.

\section{Preliminary Results}

\begin{figure*}
\begin{tabular}{cc}
\psfig{file=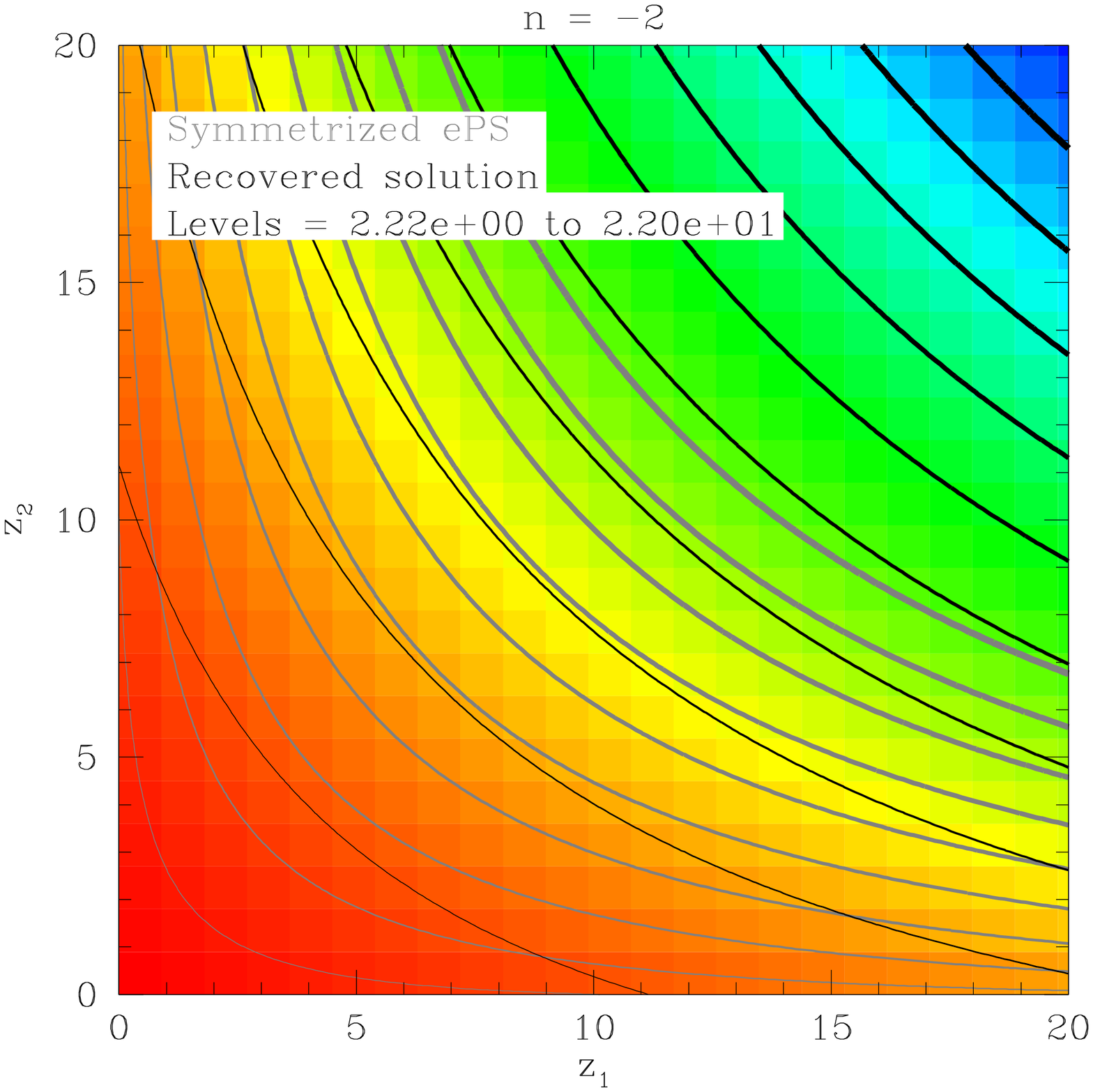,width=80mm} & \psfig{file=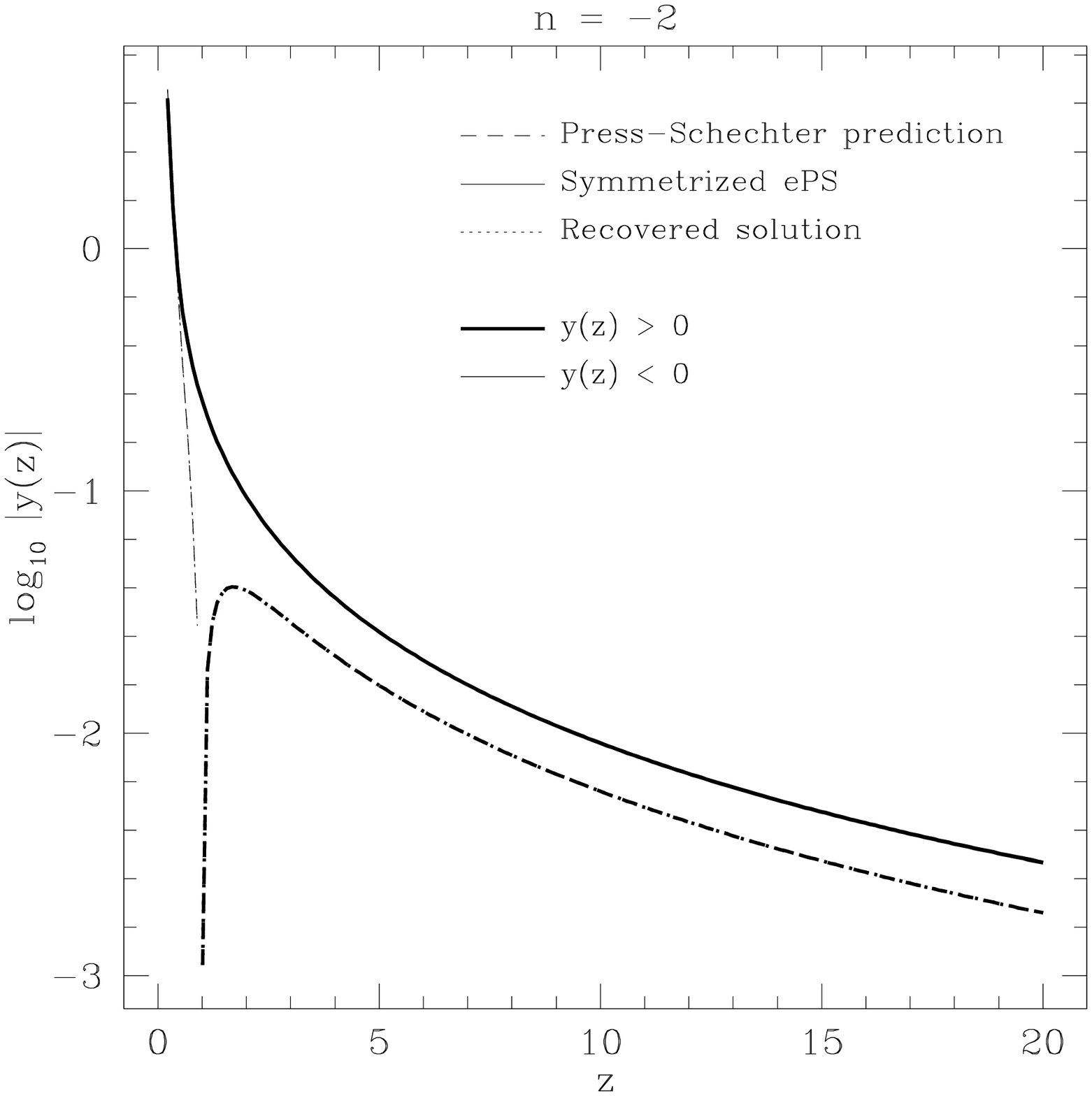,width=80mm}
\end{tabular}
\caption{\emph{Left-hand panel:} The recovered solution for
$q(z_1,z_2)$ for an $n=-2$ power spectrum is shown as a contour plot
(black lines) and by the coloured shading. For comparison, we plot the
symmetrized extended-Press-Schechter prediction as grey
contours. Contours are drawn between the values indicated in the
figure at equal logarithmic intervals (thinnest line corresponding to
lowest value). \emph{Right-hand panel:} The solution, $y(z)$, for an
$n=-2$ power spectrum to the Smoluchowski equation. The dashed black
line shows the exact Press-Schechter prediction, while the dotted line
shows the result found using the Press-Schechter mass function, along
with the numerically recovered solution for $q(z_1,z_2)$. (Note that
these two lines coincide, giving the appearance of a single dot-dashed
line.) The solid line shows the result obtained by using the
Press-Schechter mass function together with the symmetrized extended
Press-Schechter prediction for $q(z_1,z_2)$---we plot results for both
arithmetically and geometrically symmetrized $q(z_1,z_2)$, but the two
are indistinguishable in these figures. Note that we plot the absolute
value of $y(z)$ and indicate regions where the function becomes
negative by thinner lines.}
\label{fig:res_n-2}
\end{figure*}

\begin{figure*}
\begin{tabular}{cc}
\psfig{file=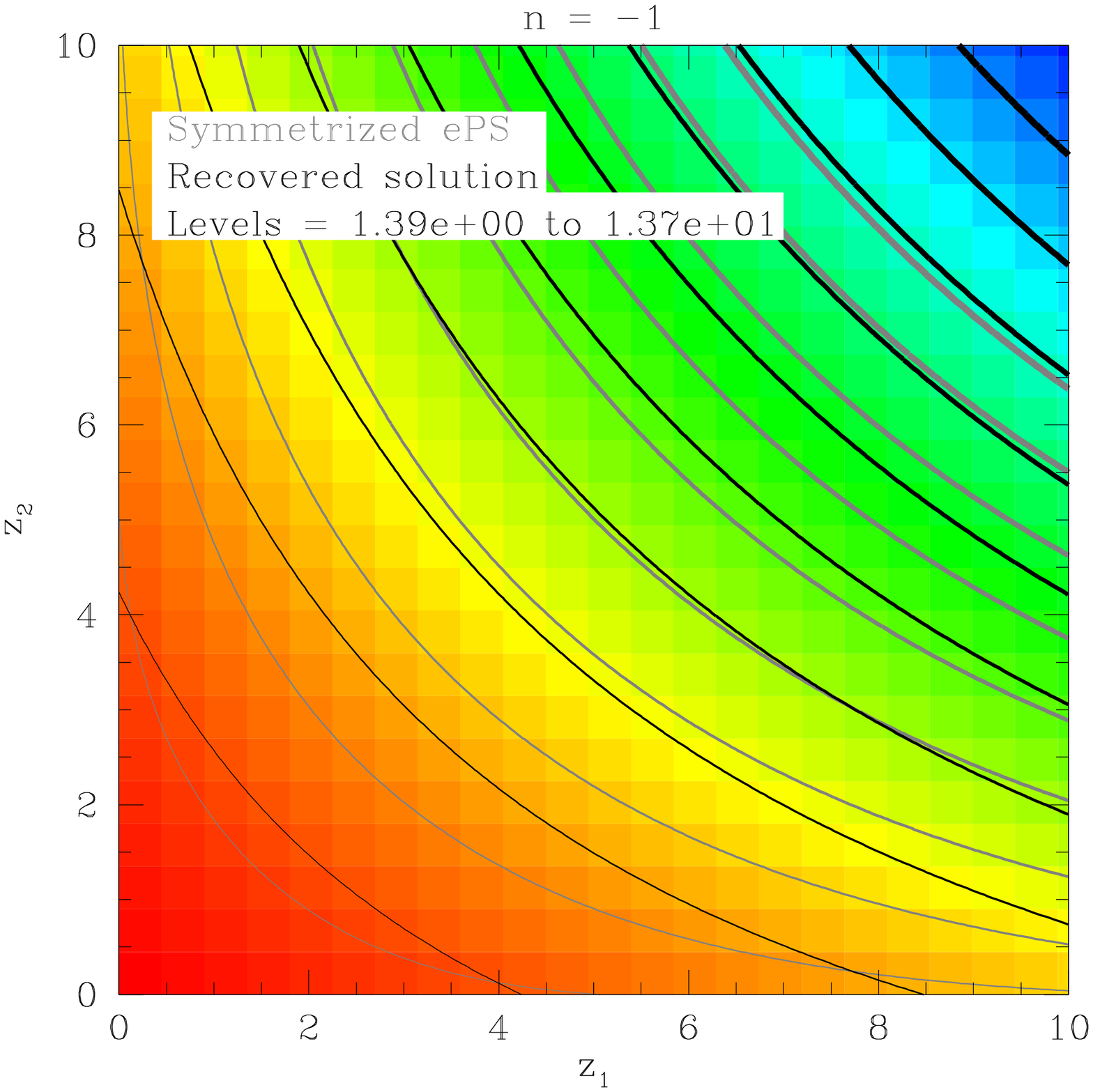,width=80mm} & \psfig{file=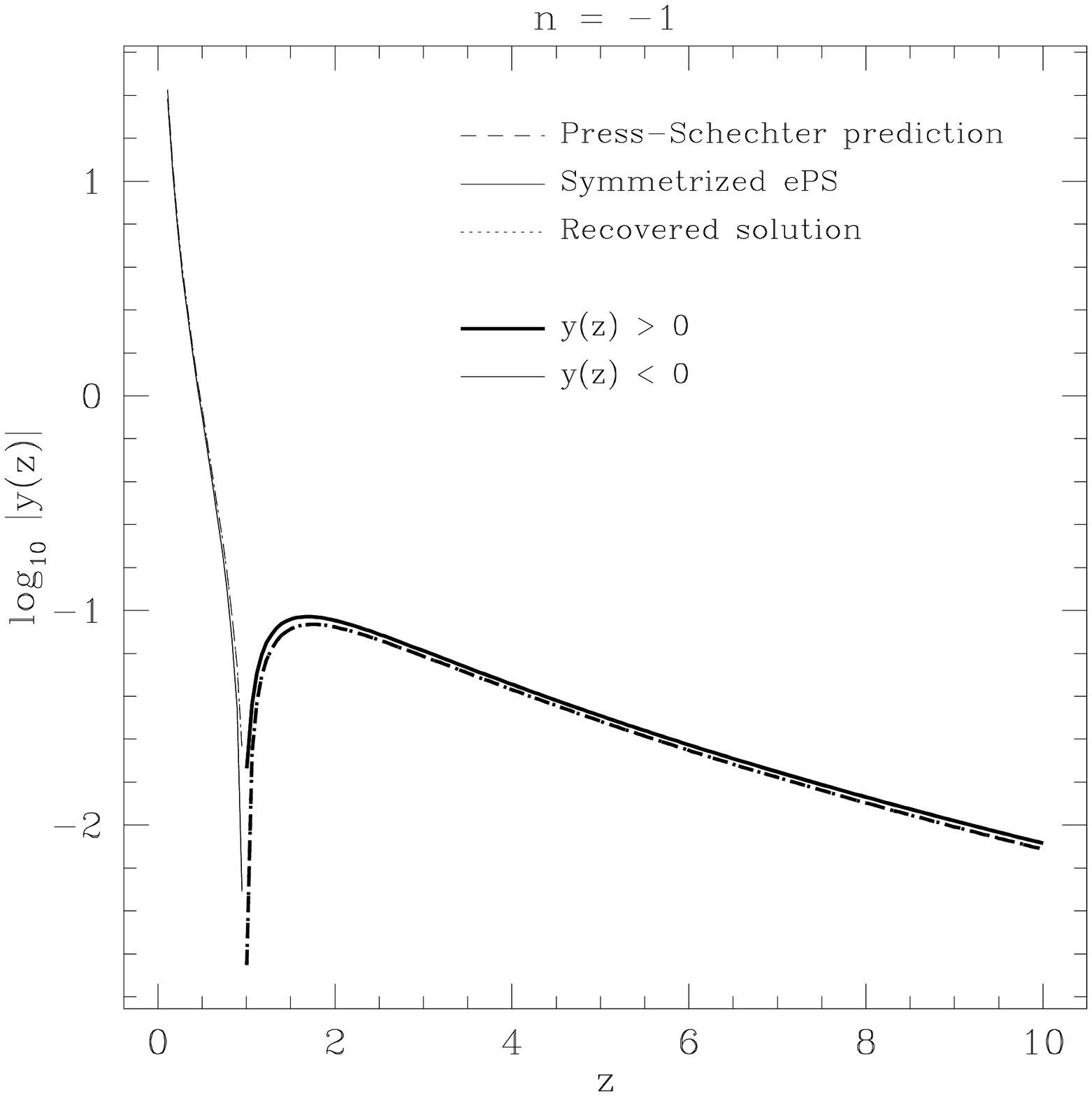,width=80mm}
\end{tabular}
\caption{As Fig.~\protect\ref{fig:res_n-2} but for $n=-1$.}
\label{fig:res_n-1}
\end{figure*}

\begin{figure*}
\begin{tabular}{cc}
\psfig{file=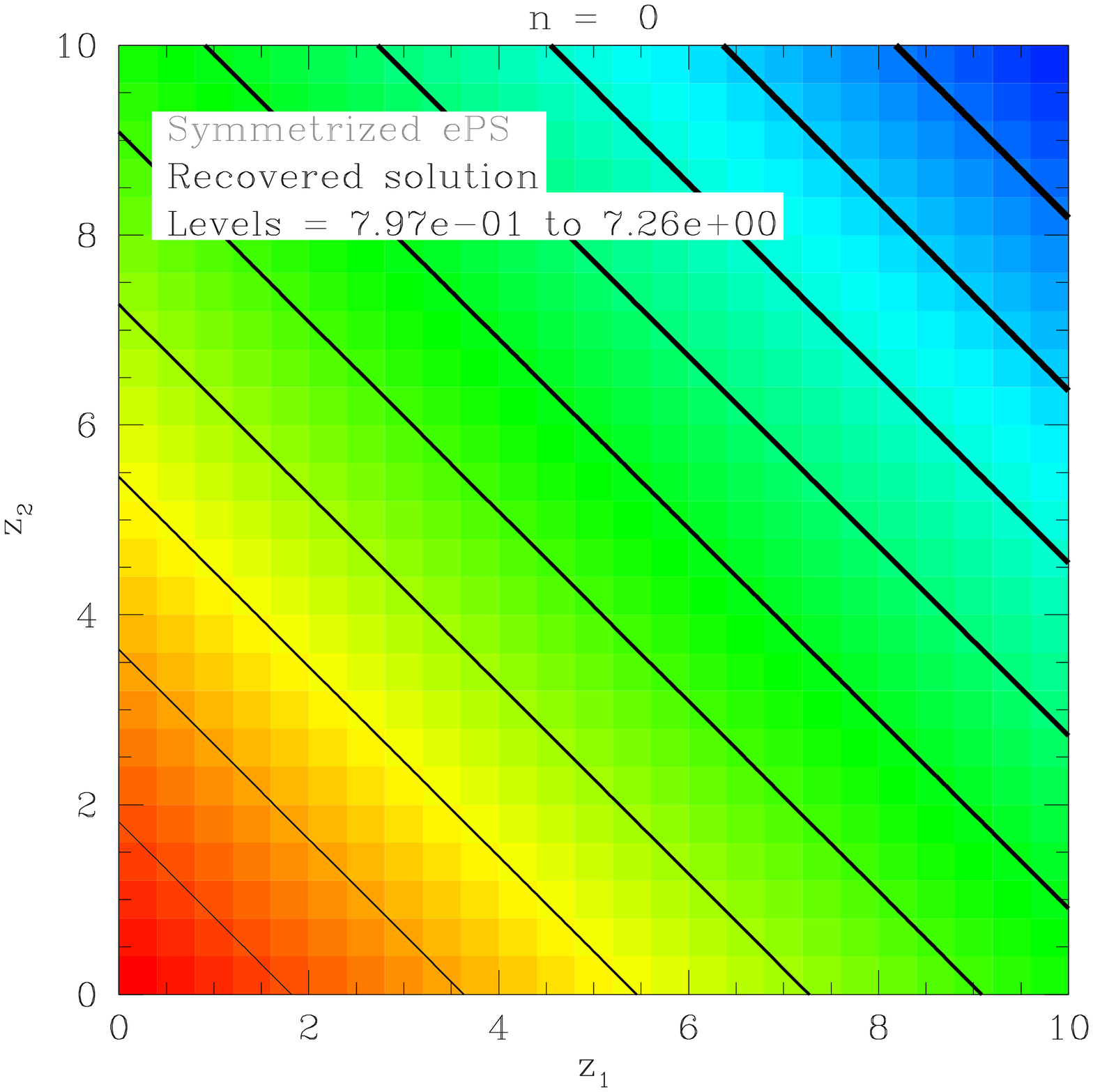,width=80mm} & \psfig{file=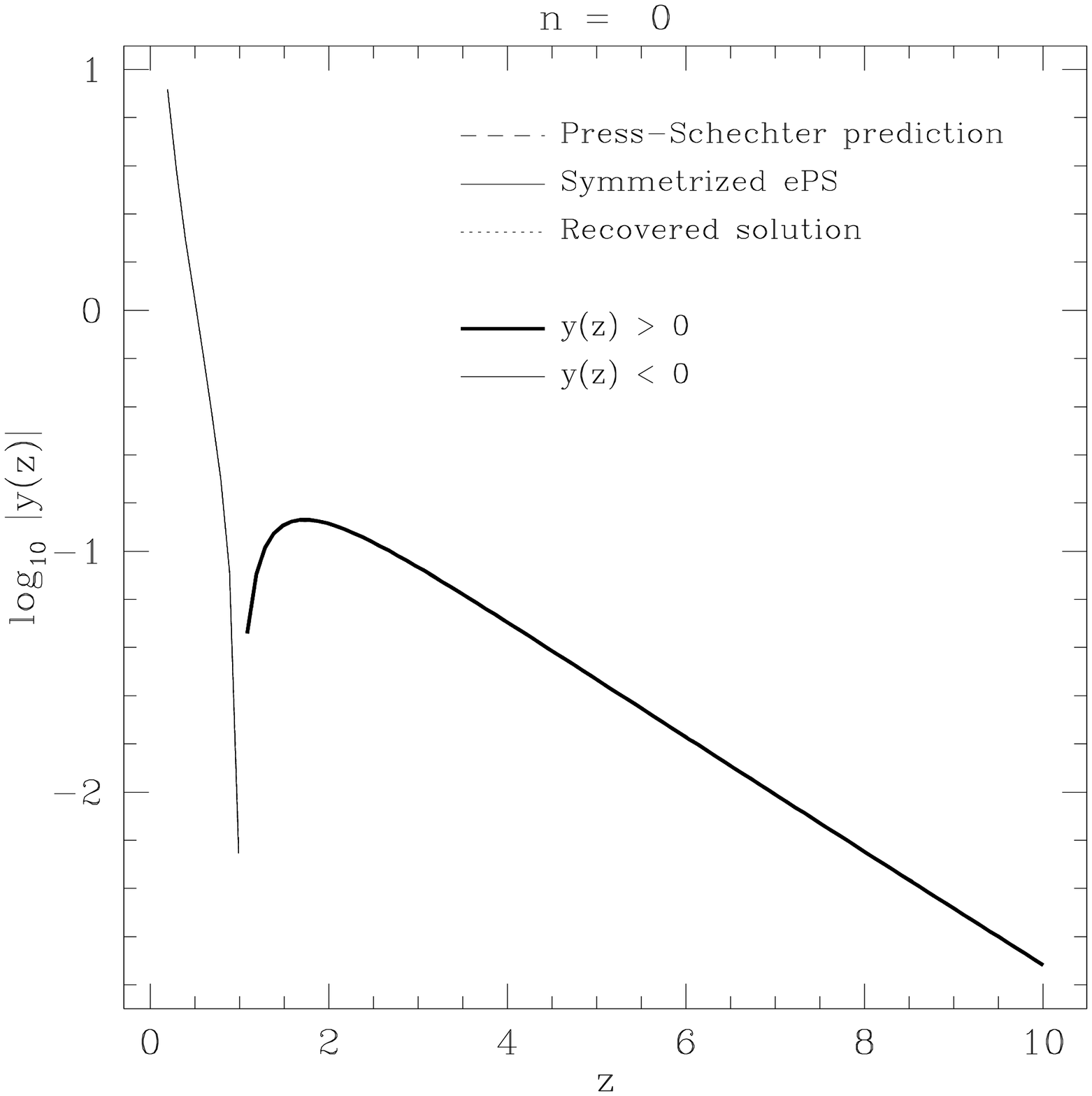,width=80mm}
\end{tabular}
\caption{As Fig.~\protect\ref{fig:res_n-2} but for $n=0$.}
\label{fig:res_n0}
\end{figure*}

\begin{figure*}
\begin{tabular}{cc}
\psfig{file=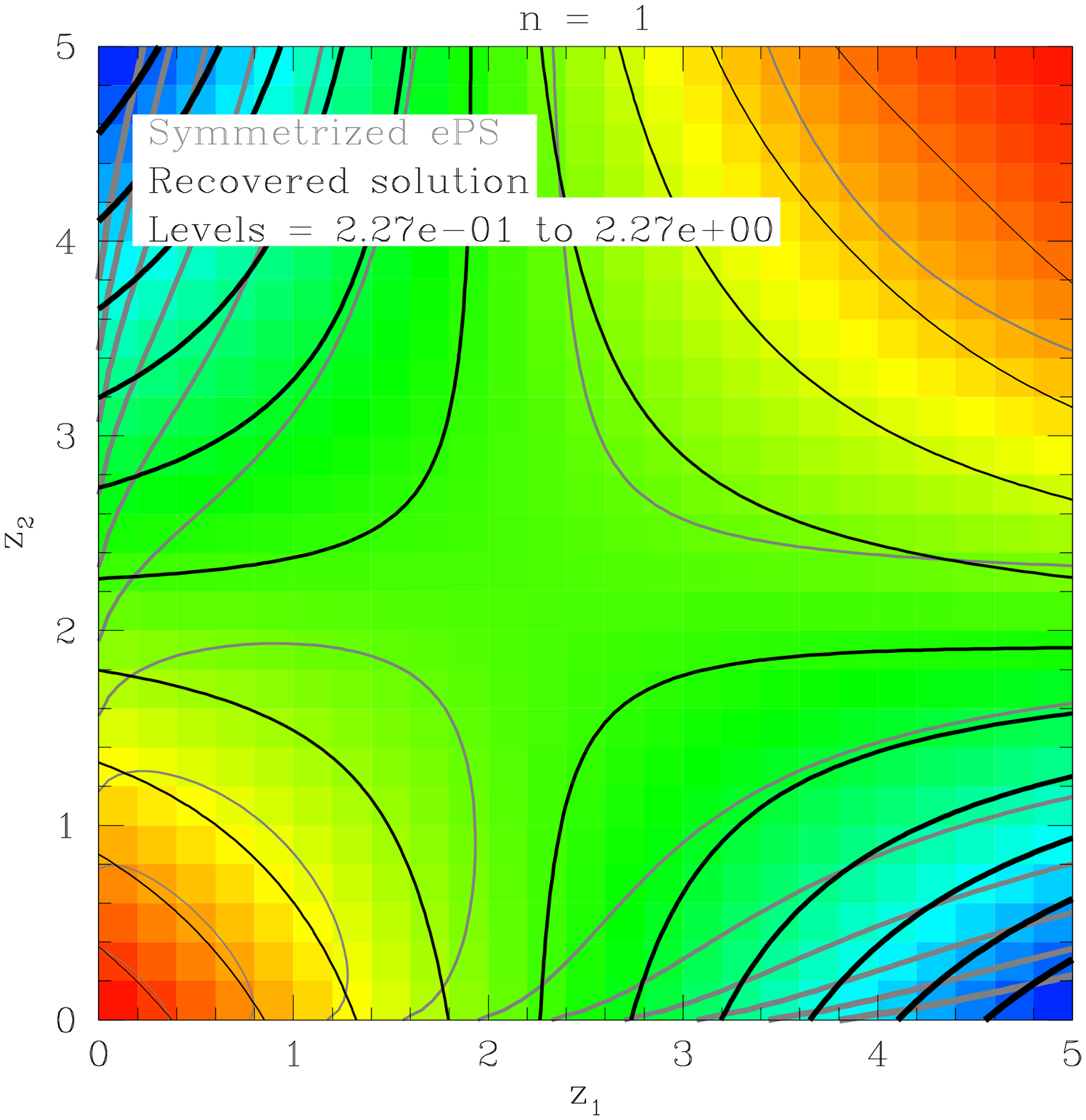,width=80mm} & \psfig{file=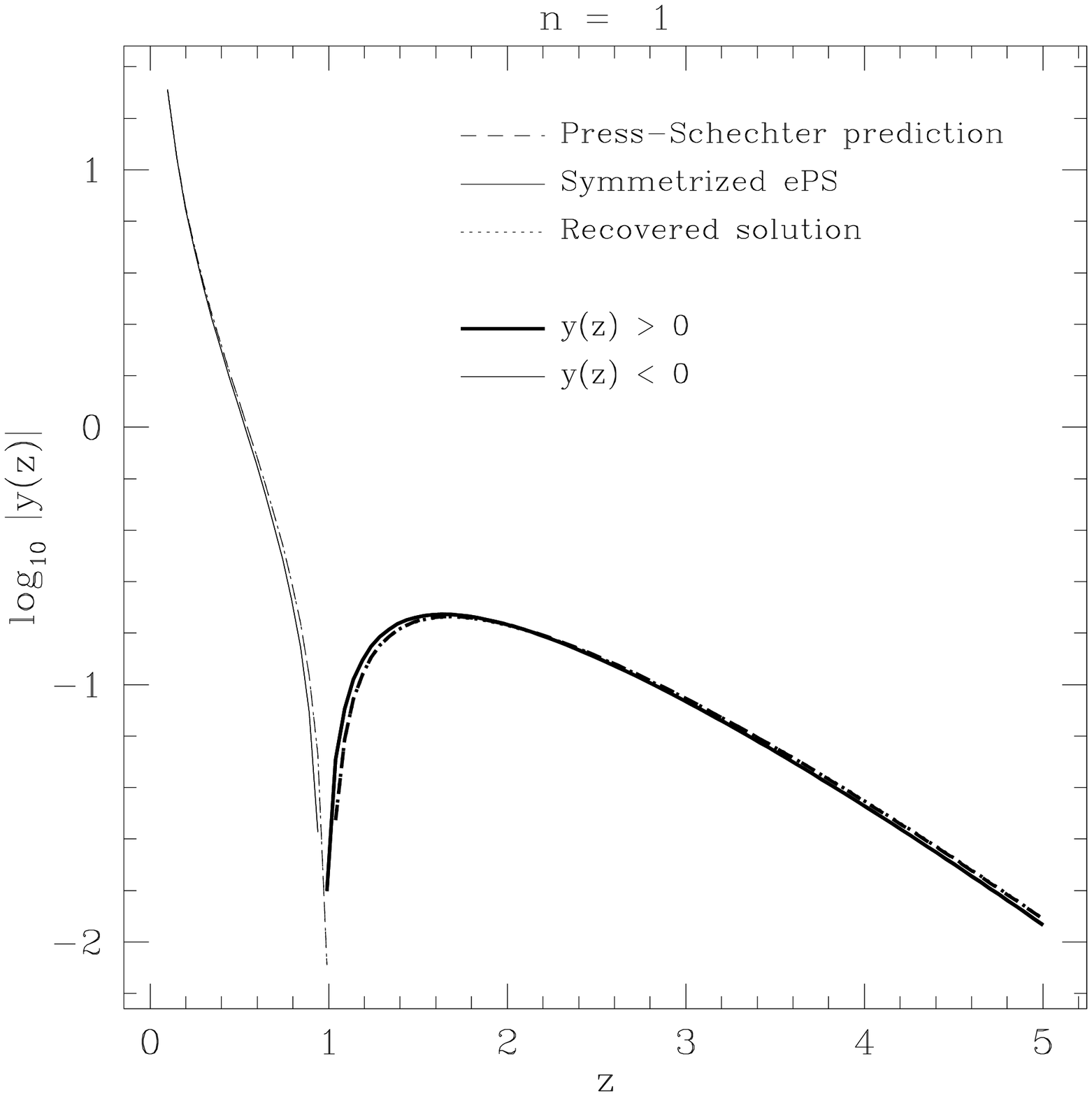,width=80mm}
\end{tabular}
\caption{As Fig.~\protect\ref{fig:res_n-2} but for $n=1$.}
\label{fig:res_n1}
\end{figure*}

\begin{figure*}
\begin{tabular}{cc}
\psfig{file=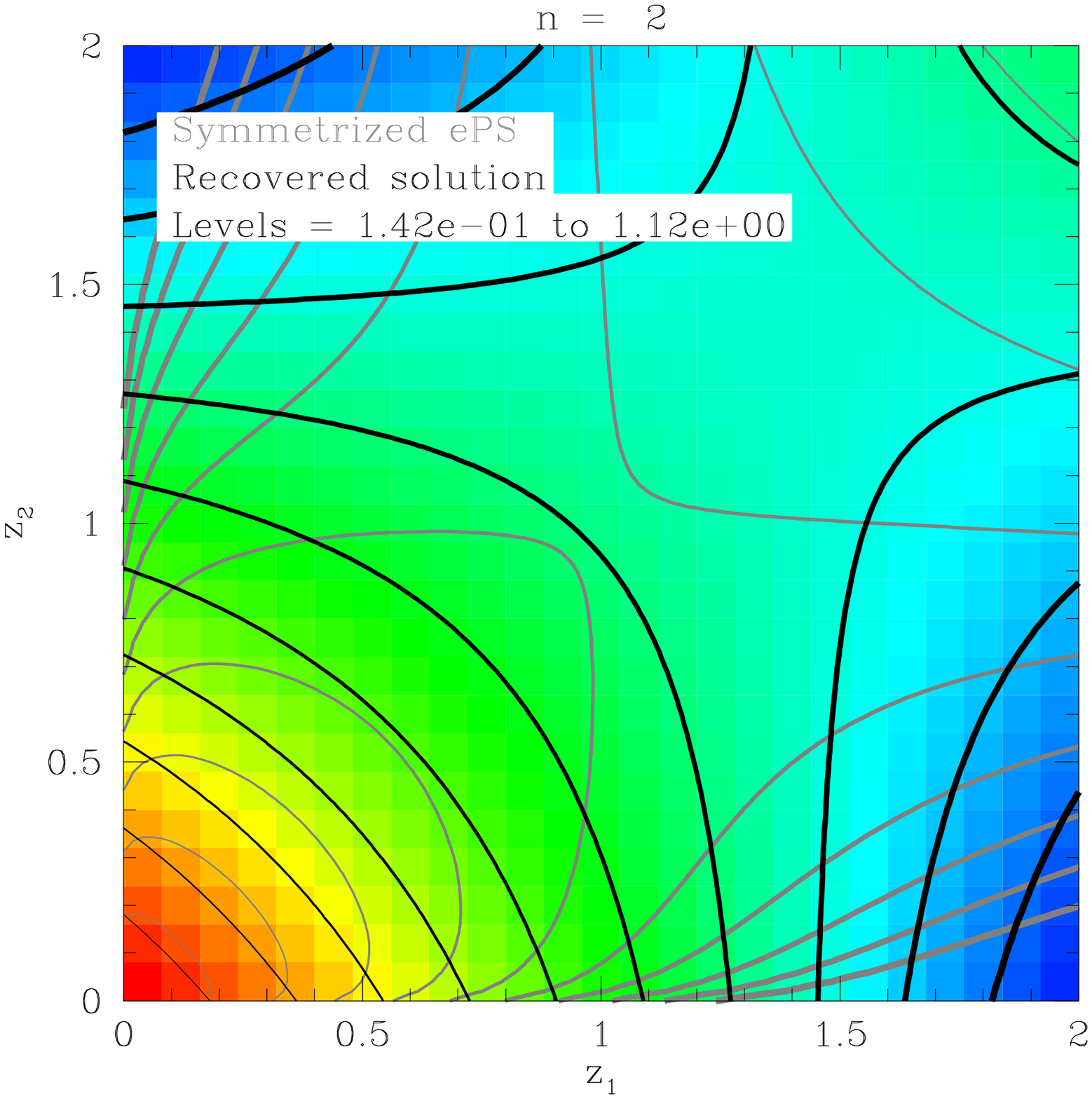,width=80mm} & \psfig{file=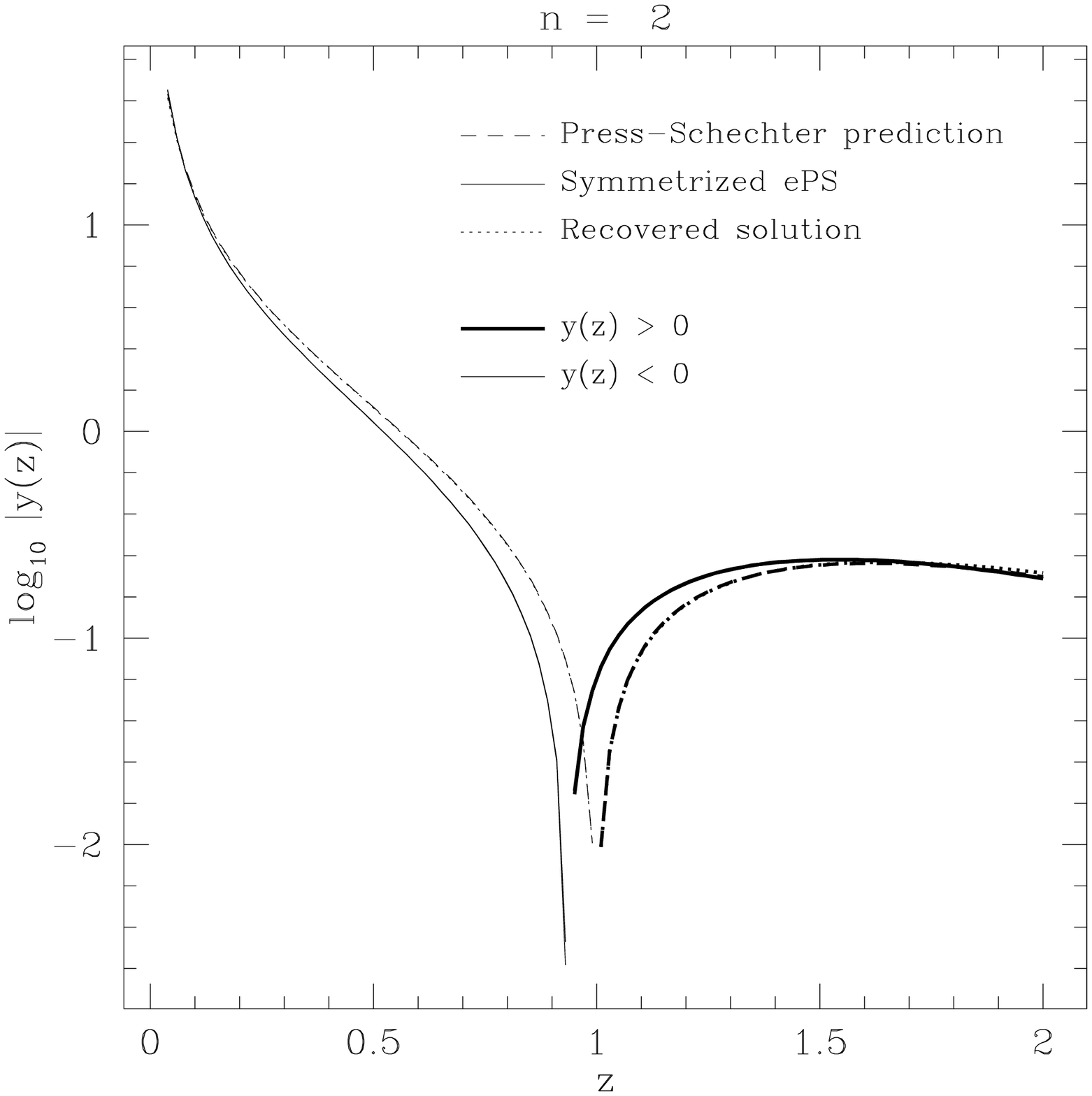,width=80mm}
\end{tabular}
\caption{As Fig.~\protect\ref{fig:res_n-2} but for $n=2$.}
\label{fig:res_n2}
\end{figure*}

\begin{figure*}
\begin{tabular}{cc}
\psfig{file=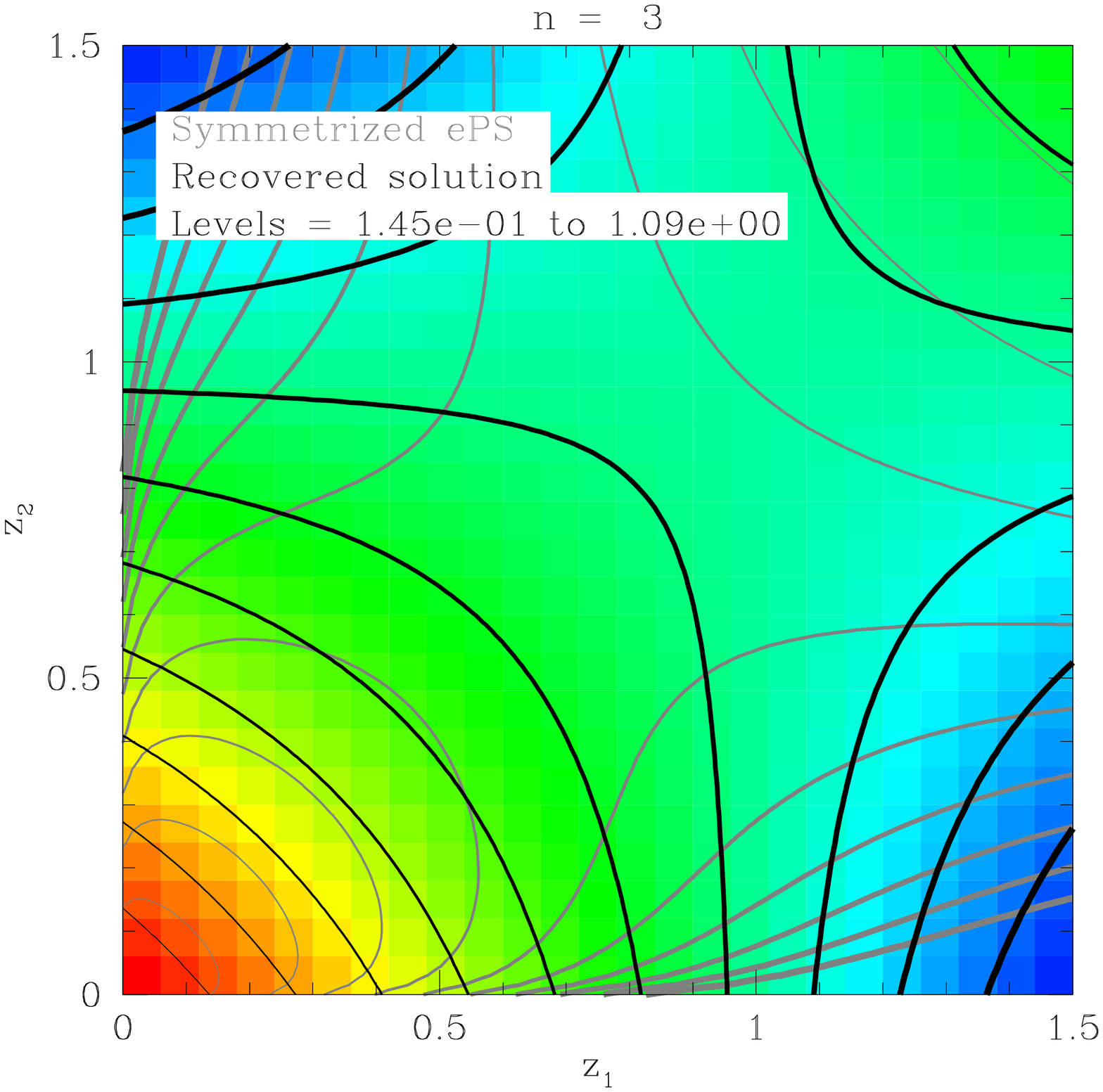,width=80mm} & \psfig{file=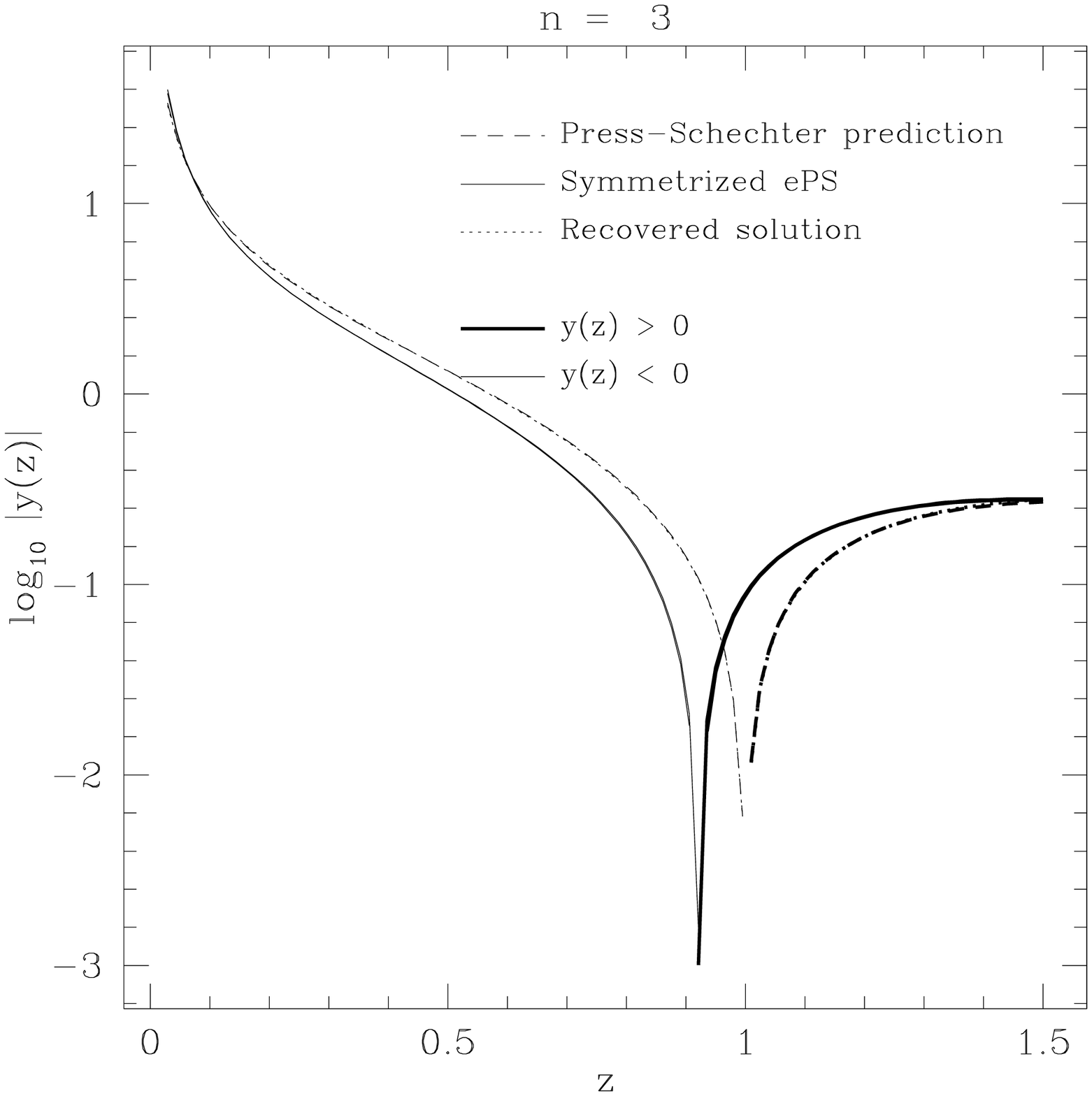,width=80mm}
\end{tabular}
\caption{As Fig.~\protect\ref{fig:res_n-2} but for $n=3$.}
\label{fig:res_n3}
\end{figure*}

Calculations have been performed for power-law power spectra with
$n=-2$, $-1$, $0$, $1$, $2$ and $3$. For the $n=0$ case an exact
solution is known,
$q(z_1,z_2)=(z_1+z_2)/\sqrt{2\pi}$. Figures~\ref{fig:res_n-2} through
\ref{fig:res_n3} show, in their left-hand panels, contour maps of the
function $q(z_1,z_2)$ recovered by the solution method described above
for each value of $n$. The functions are clearly symmetric in their
arguments and are all smoothly varying. For contrast, grey lines show
a geometrically symmetrized extended Press-Schechter prediction
$q_{\rm ePS,sym,G}(z_1,z_2) = \left[ q_{\rm ePS}(z_1,z_2)q_{\rm
ePS}(z_2,z_1) \right]^{1/2}$, where $q_{\rm ePS}$ is the extended
Press-Schechter merger rate corresponding to
equation~(\ref{eq:qeps}). The right-hand panels of these Figures show
$y(z)$ predicted by Press-Schechter theory, together with the $y(z)$
determined from the Smoluchowski equation using ${\bf q}$ determined
by the techniques described above and using the arithmetically
($q_{\rm ePS,sym,A}(z_1,z_2) = \left[ q_{\rm ePS}(z_1,z_2) + q_{\rm
ePS}(z_2,z_1) \right]/2$) and geometrically symmetrized extended
Press-Schechter kernels. (Note that the results for arithmetically and
geometrically symmetrized extended Press-Schechter kernels are
indistinguishable in these figures.) In every case we are able to find
a symmetric, smoothly varying solution which solves the Smoluchowski
equation. The solutions typically differ significantly from the
symmetrized extended Press-Schechter prediction, which does not solve
the Smoluchowski equation (except for the specific case of $n=0$).

The results obtained depend upon several factors:
\begin{enumerate}
\item The relative importance given to solving the Smoluchowski
equation and meeting the regularization condition when searching for a
solution (i.e. the value of $\sigma_{R_1}$).
\item The number of elements used in the discretized representation of
the Smoluchowski equation and the range of $z$ studied (i.e. $N$ and
$\Delta z$). (We have adopted the approach of making $N$ as large as
possible given constraints on computing resources, and to make $\Delta
z$ as large as possible while retaining sufficient resolution to find
an accurate solution to the Smoluchowski equation.)
\item The nature of the regularization condition(s).
\end{enumerate}
The first and second factors can be addressed easily given sufficient
computing resources\footnote{Our current calculations represent the
function $q(z_1,z_2)$ by an $N\times N$ grid, with $N=179$. Accounting
for the symmetric nature of $q(z_1,z_2)$ this requires $N(N+1)/2$
elements. The matrix to be inverted therefore contains $N^2(N+1)^2/4$
elements, so memory requirements scale as $N^4$. Given that matrix
inversion is an $N^3$ process (or, at best, $N^{2.807}$) the time
required to find a solution scales as $N^6$. Thus, merely doubling $N$
increases memory requirements by a factor of 16 and time requirements
by a factor of 64. With $N=179$ memory requirements are of order 2~Gb,
while finding a solution takes around $3\times 10^5$ seconds on a
3~GHz workstation. The choice $N=179$ was therefore dictated by both
being able to fit the calculation into the memory of a 4~Gb
workstation and taking a not unreasonable amount of time to
run. Furthermore, the compiler used has a maximum array size which
prevents us from making the matrix any larger, although this could be
trivially circumvented by using multiple arrays to represent our
matrices.}, while the third may require consideration of other
constraints that any solution which is to be considered physically
reasonable must meet. With the current maximum value of $N=179$ we
find that our results change at the 10--20\% level if we reduce $N$ to
$101$, while changes in $\Delta z$ at fixed $N$ similarly lead to
changes in $q(z_1,z_2)$ at the 10--20\% level.
Clearly there is a need to increase $N$ further, or find ways to
perform the inversion of the Smoluchowski equation more robustly. We
find that our results are quite insensitive to $\sigma_{R_1}$
providing it is not made too large or too small (i.e. there is a wide
range of $\sigma_{R_1}$ over which the results do not change
significantly). It seems that the simple condition that the function
be smoothly varying is sufficient to obtain physically plausible
solutions.

Our results can be described moderately well by the following fitting
formula:
\begin{eqnarray}
     q(z_1,z_2) &=& {1 \over \sqrt{2\pi}} \left(A_1[z_1+z_2] +
     A_2 z_1 z_2  \right) \nonumber \\
     & & \times \exp\left( -A_4 [z_1 z_2]^{A_3} \right), 
\label{eq:fit}
\end{eqnarray}
where $A_1$, $A_2$, $A_3$ and $A_4$ are free parameters which we
determine by fitting to the numerically determined
$q(z_1,z_2)$. Table~\ref{tb:fitpar} lists the values of these
parameters. The final column of that table lists the maximum
percentage deviation between the fitting formula and the numerical
results as shown in Figures~\ref{fig:res_n-2} through
\ref{fig:res_n3}. These fitting formulae are valid over the range of
$z$ shown in the figures, but cannot be guaranteed to hold for $z$'s
outside of these ranges. In particular, for $n>0$ the parameter $a_2$
is negative, which will result in $q(z_1,z_2)<0$ for certain
$z_1,z_2$, which is clearly unphysical. We hope to establish a better
fitting formula in future work.
 
\begin{table}
\caption{Values for the parameters,f $A_1$ to $A_4$, for the fitting
formula given in equation~(\protect\ref{eq:fit}) as a function of
power-spectrum index, $n$. The final column lists the maximum
percentage difference between the fitting formula and the numerical
results, when using the values given in the table, over the range of
$z_1$ and $z_2$ shown in Figures~\protect\ref{fig:res_n-2} through
\protect\ref{fig:res_n3}.}
\label{tb:fitpar}
\begin{tabular}{cccccc}
\hline
$n$ & $A_1$ & $A_2$ & $A_3$ & $A_4$ & Max. \% deviation \\
\hline
-2 & 0.502 &  0.102 & -0.450 &  0.118 & 15.0 \\
-1 & 0.809 &  0.219 & -0.504 &  0.027 & 8.2 \\
0  & 1.000 &  0.000 &  0.000 &  0.000 & 0.0 \\
1  & 1.239 & -0.496 &  1.580 &  0.009 & 14.7 \\
2  & 1.547 & -1.089 & -0.555 & -0.004 & 4.9 \\
3  & 1.997 & -2.000 & -0.561 & -0.005 & 9.3 \\
\hline
\end{tabular}
\end{table}

\begin{figure*}
\begin{tabular}{cc}
\psfig{file=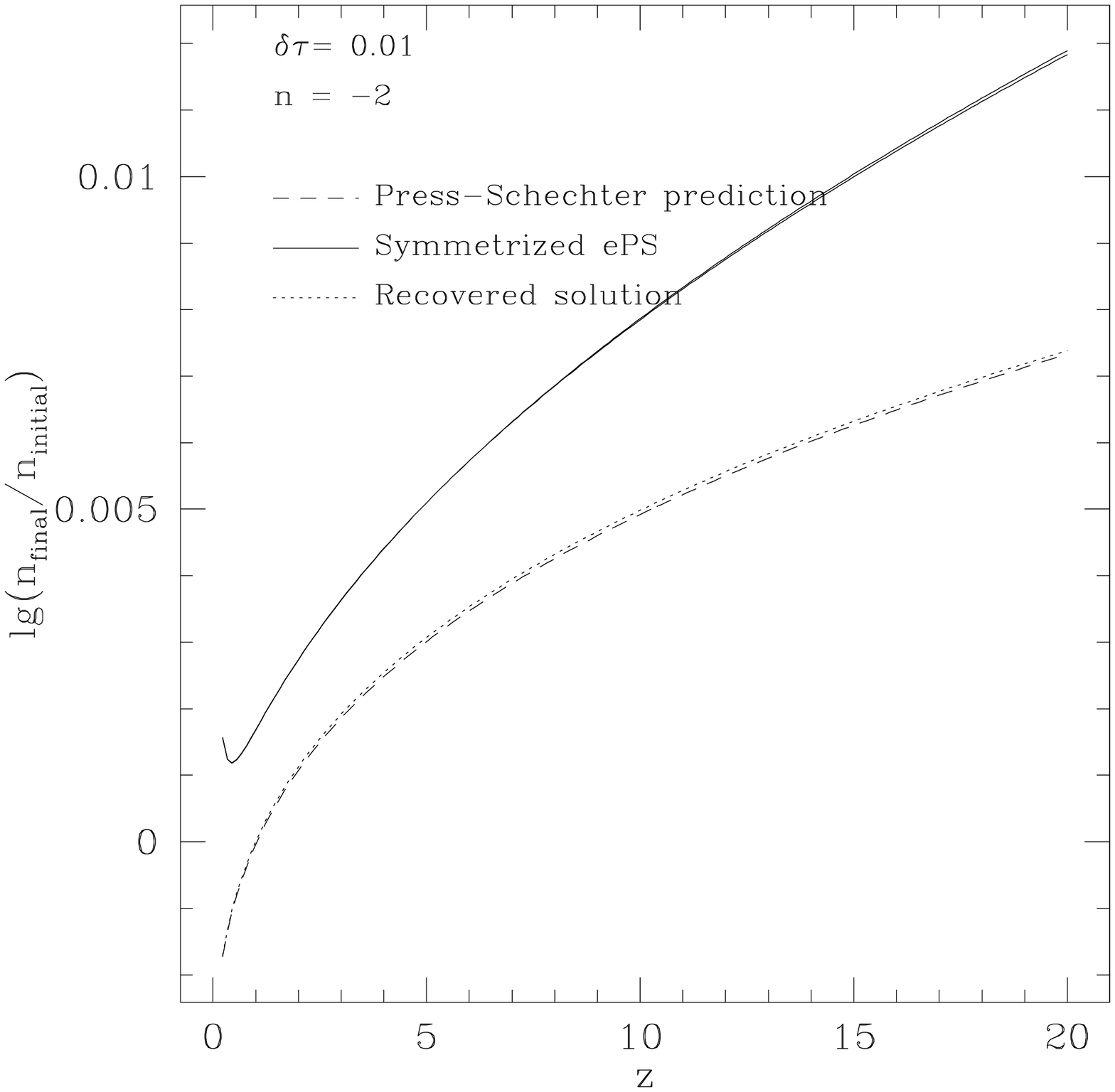,width=70mm} & \psfig{file=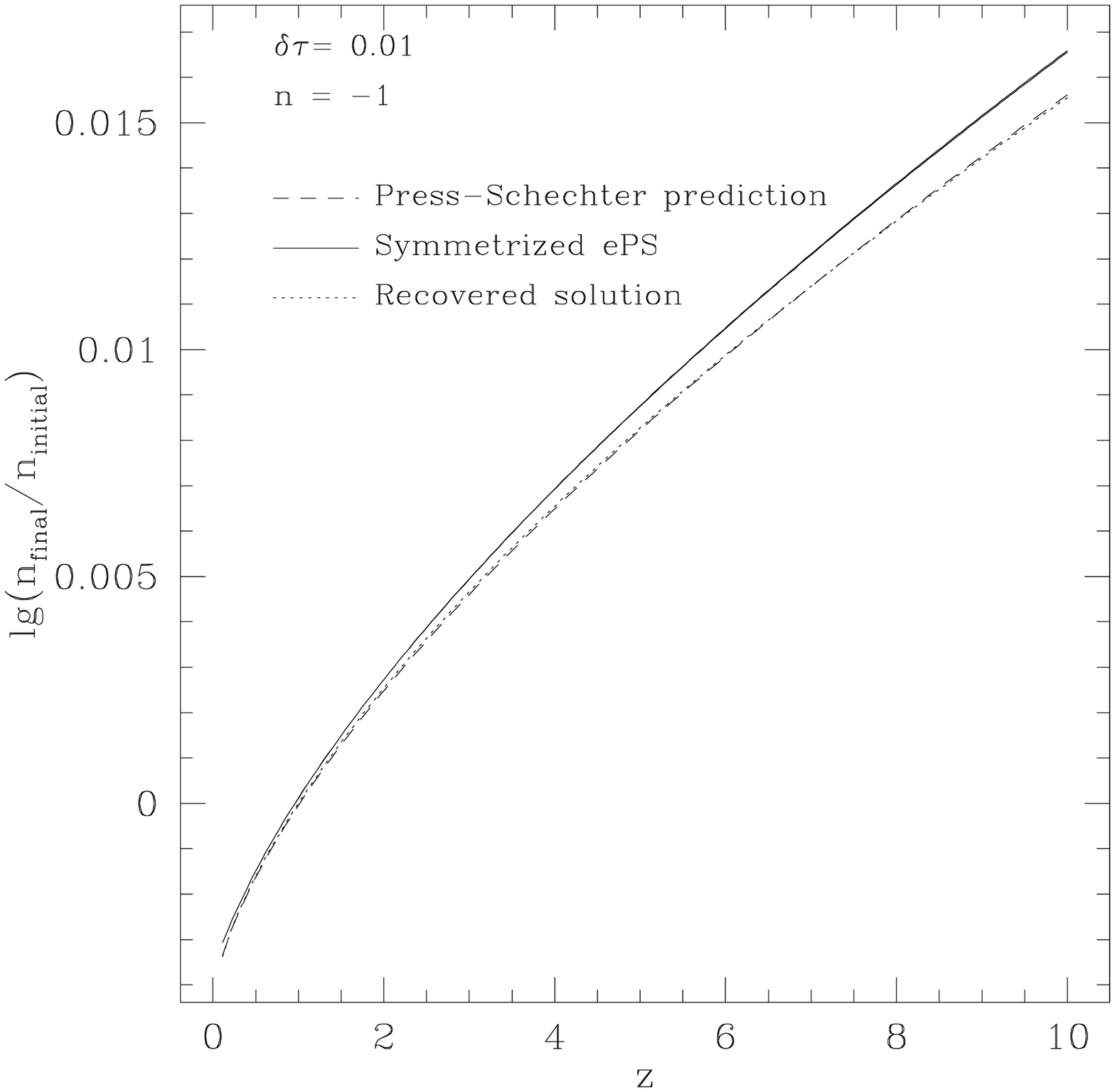,width=70mm} \\
\psfig{file=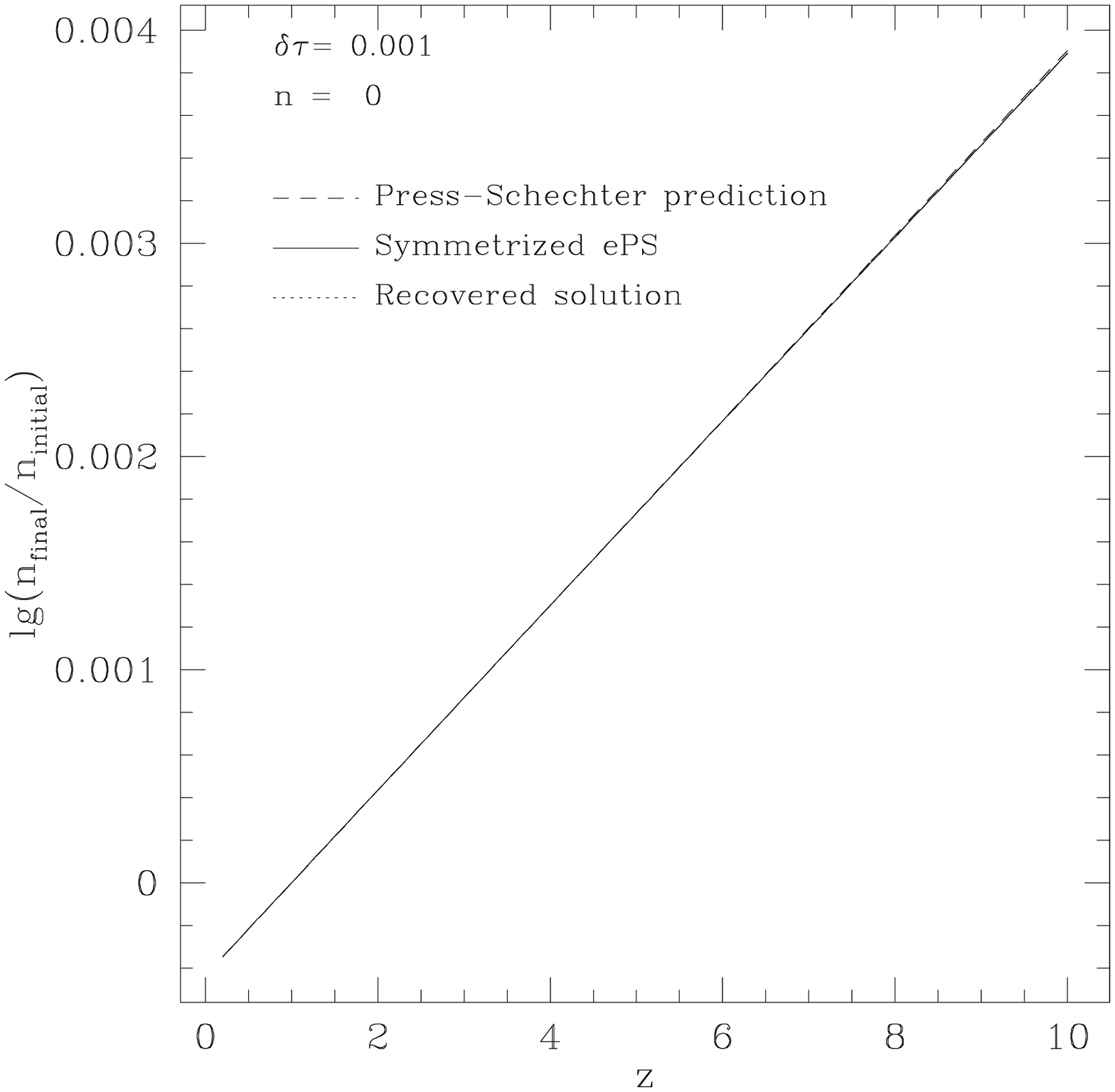,width=70mm} & \psfig{file=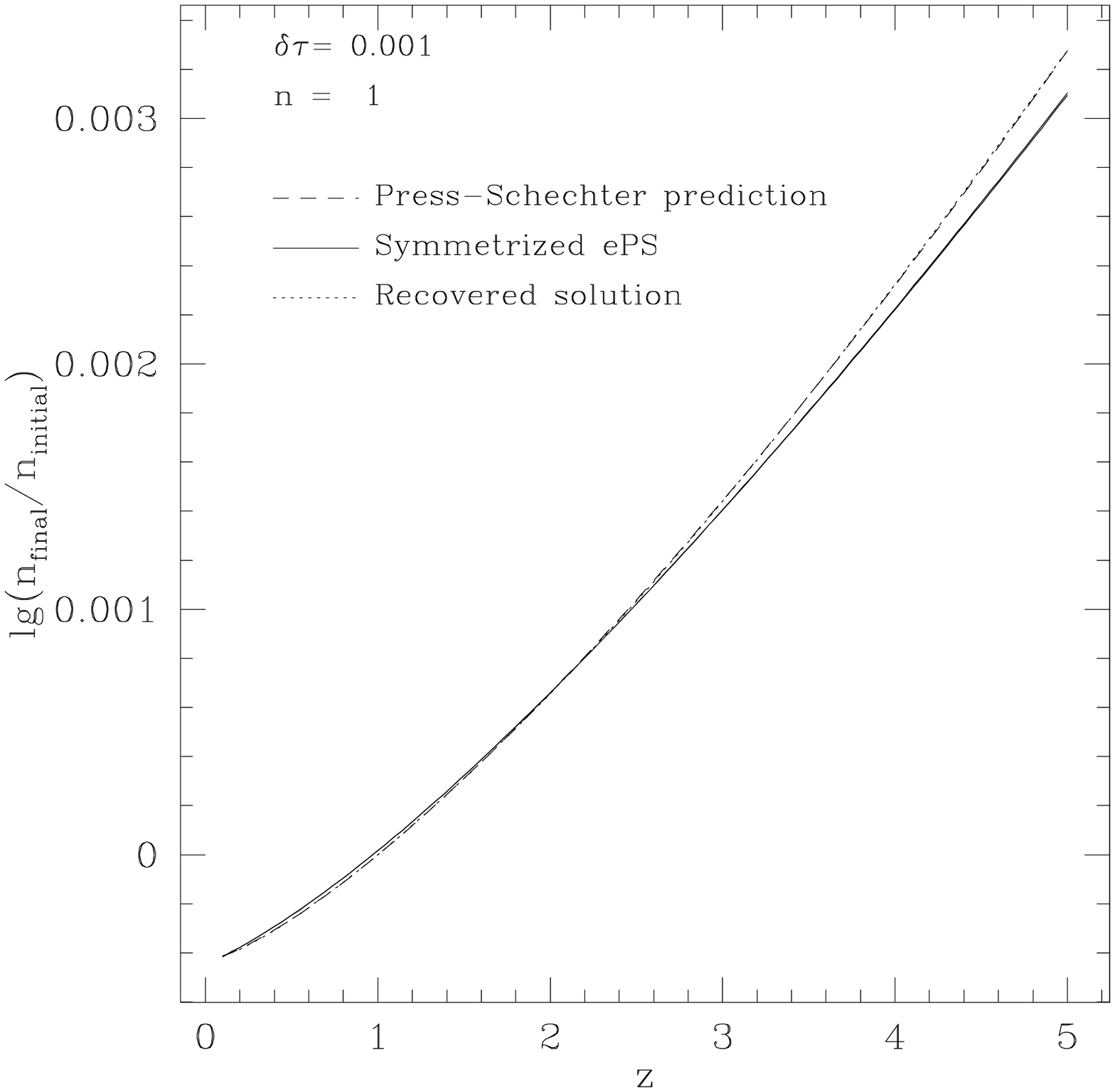,width=70mm} \\
\psfig{file=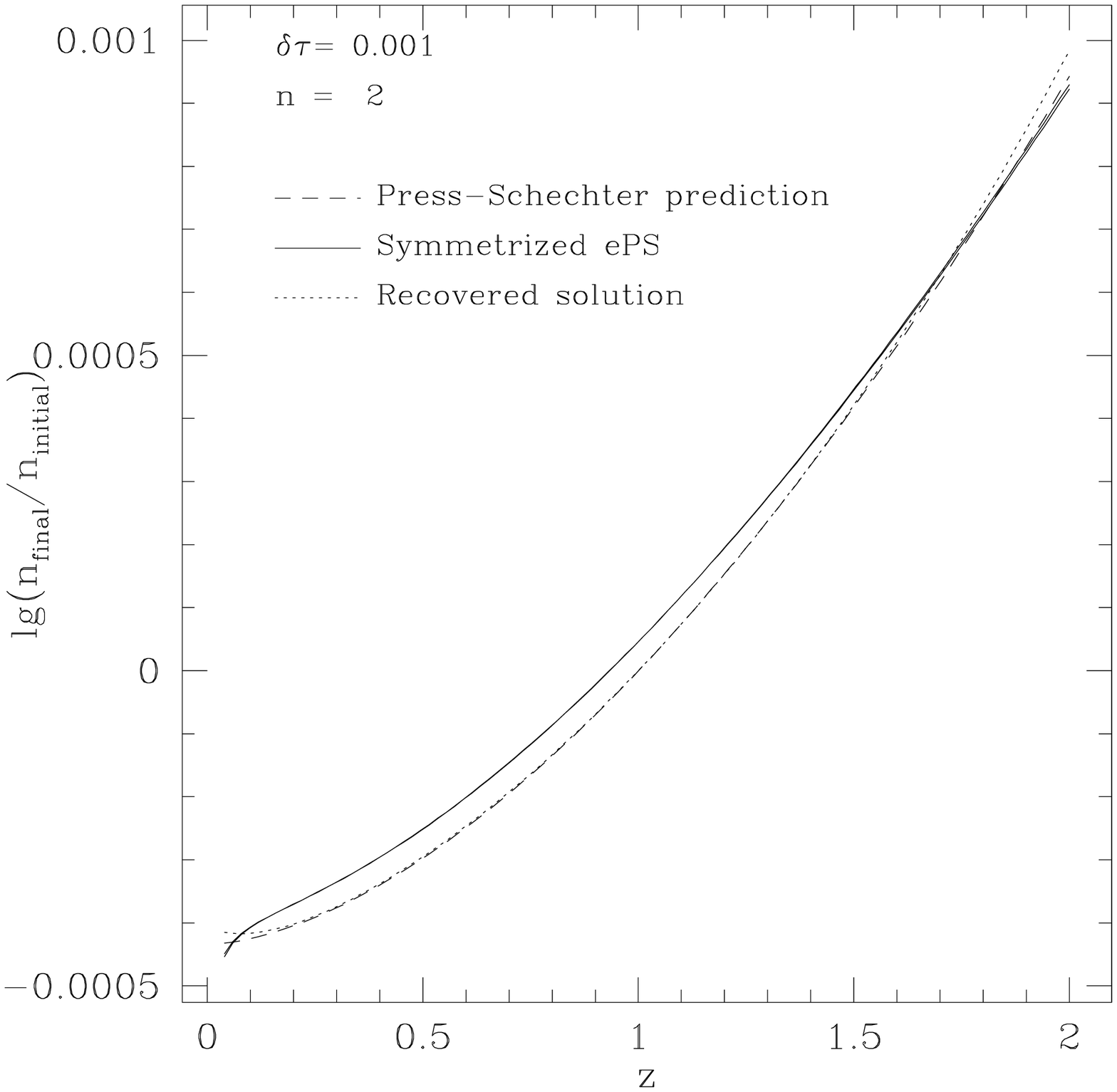,width=70mm} & \psfig{file=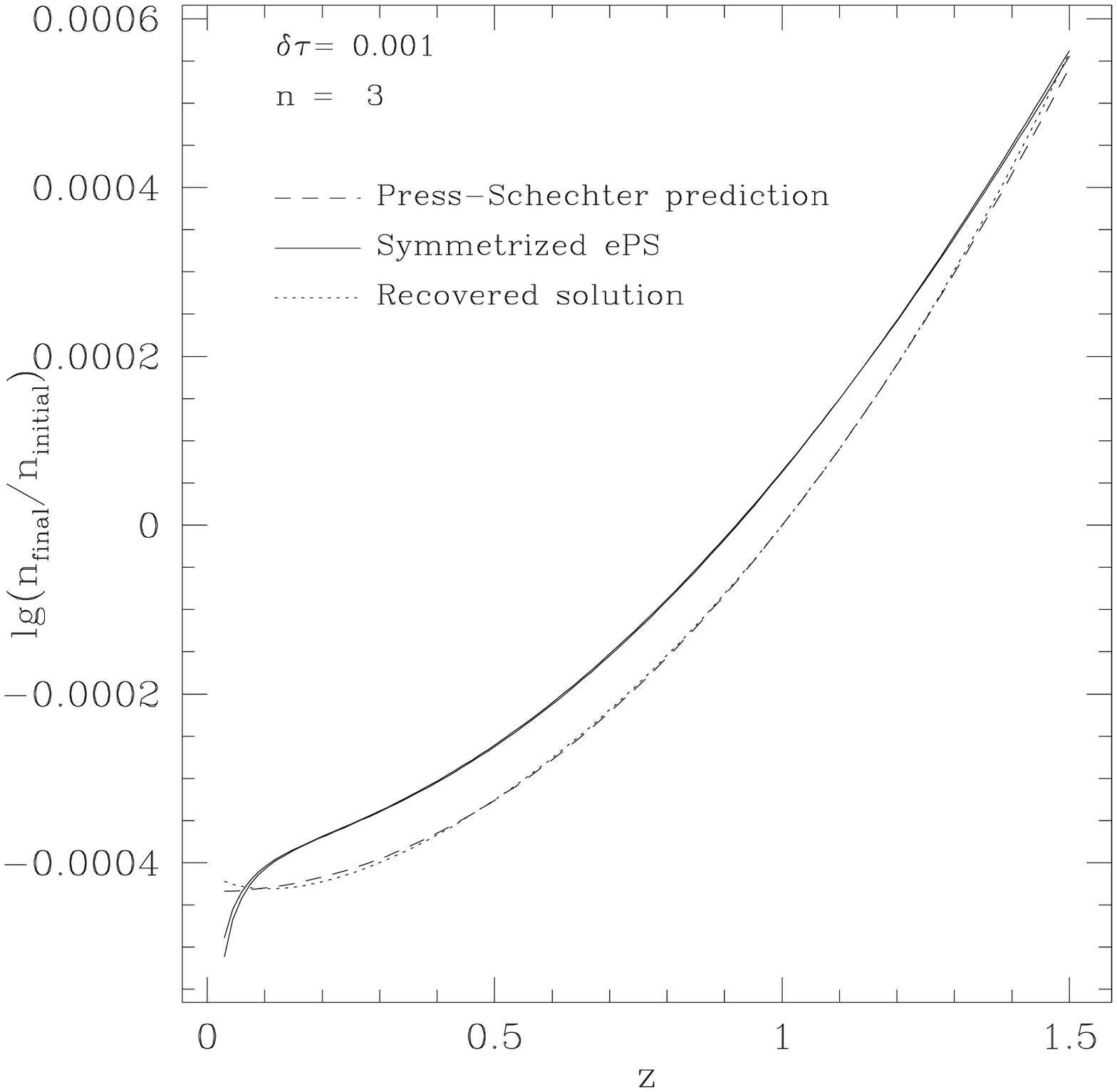,width=70mm} \\
\end{tabular}
\caption{The evolution of the Press-Schechter mass function over a
small time step $\delta\tau$ (the value of which is given in each
panel). We plot the ratio of the final to initial mass function as a
function of mass, $z$. The dashed line shows the expectation from the
Press-Schechter theory, while the dotted line shows the result of
applying our recovered merger rates to the initial mass function. The
solid lines show the results of applying the two symmetrized extended
Press-Schechter merger rates to the initial mass function (these two
lines are indistinguishable almost everywhere in this figure).}
\label{fig:step}
\end{figure*}

It is clear from the right-hand panels of Figs.~\ref{fig:res_n-2}
through \ref{fig:res_n3} that our solutions for $q(z_1,z_2)$ produce
the correct rate of change of halo abundance. To confirm this we can
consider the distribution of halo masses at time $\tau$, $n(z;\tau)$,
where $z=M/M_*(\tau)$. At time $\tau+\delta \tau$ the distribution of
halo masses, using the same definition of $z=M/M_*(\tau)$ and
\emph{not} $z=M/M_*(\tau+\delta\tau)$, is
$n(z;\tau+\delta\tau)=f[\delta\tau]^2n(f[\delta\tau]z;\tau)$ where
$f(x)=\exp[-6x/(3+n)]$. Using our recovered $q(z_1,z_2)$ we can evolve
$n(z;\tau)$ to $n(z;\tau+\delta\tau)$ for a small step $\delta\tau$
(such as would be used in a merger tree calculation). We show the
results of this test in Fig.~\ref{fig:step}, which demonstrates our
ability to reproduce the evolution of the Press-Schechter mass
function using our merger rates. (The figure also demonstrates how
poorly the symmetrized extended Press-Schechter estimate of the
merging rate does.)

\subsection{Some comments on the numerical results}

The numerical and ePS merger kernels appear to be monotonically
increasing functions of mass for power-law indices $n\leq0$.
However, for power-law indices $n>0$, they appear to peak at
masses near the characteristic mass scale $M_*$.  As expected,
the agreement between the numerical and ePS results is exact for
$n=0$, and they become increasingly discrepant as $n$ departs
from 0.  In particular, our numerically-recovered merger kernel
disagrees with the ePS merger kernel, and the disagreement
increases as $n$ departs from 0.  In fact, we attempted to
find a numerical merger kernel, demanding that the merger kernel
be equal to the ePS merger kernel for equal-mass mergers.
However, our algorithm was unable to find a consistent and
smooth merger kernel with this constraint.  This leads us to
believe that the ePS merger rate is invalid, even for equal-mass
mergers where the two ePS results agree.

Finally, we point out that \scite{MurRam86} outline a numerical
technique for finding merger kernels, but only under the
assumption that the merger kernel is homogeneous; i.e., that the
merger kernel can be written in the form $q(z_1,z_2) = z_1^m
f(z_2/z_1)$, where $m$ is a power-law exponent.  We found that
we were able to implement this algorithm and recover the merger
kernel for the $n=0$ power spectrum (for which the merger kernel
is indeed homogeneous).  We then speculated that for more
general power-law power spectra, the merger kernel would also be
homogeneous.  However, we were unable to get the
Muralidar-Ramkrishna algorithm to return smooth kernels for
these power spectra.  Indeed, our numerically recovered merger
kernels are {\it not} homogeneous for these more general power
spectra. 

\section{Frequently Asked Questions}

Below we attempt to answer several questions which frequently arise in
connection with this work.

{\it What about three-body encounters?}  The assumption that halos
grow through binary mergers, which is implicit in the Smoluchowski
equation, could be questioned. Since the halo abundance diverges at
low masses it is possible that a halo will experience an infinite
number of mergers per unit time with halos of vanishingly small
mass. However, as shown in Appendix~\ref{ap:alternate}, the creation
and destruction rates due to these halos cancel exactly. Therefore, we
can truncate the halo mass function at some low mass and obtain a good
approximation to the true merging history.  Put another way, if the
mass distribution is described by discrete objects, as in an N-body
simulation, rather than a continuum mass distribution, then there are
no three-body mergers.  In such a simulation, there could easily be
three or more halos that merge within a finite time interval.
However, one can always find a sufficiently small time step so that
any apparent three-body merger will in fact be a rapid sequence of
two-body mergers (somewhat like the triple-alpha reaction in stellar
nucleosynthesis, which is in fact a rapid series of two-body
reactions).

{\it What can we learn from N-body simulations?}  In principle, the
merger kernel can be determined from N-body simulations.  In practice,
though, this will be challenging. If we have $N$ mass bins, then there
will be $N(N+1)/2$ entries in the merger kernel, and we must have a
sufficiently large simulation to provide a statistically significant
number of mergers in each of these bins.  Such a simulation will also
require a very large dynamic range so that a huge number of halos of
widely varying masses will be resolved.  However, there may be a
simpler avenue. If we carry out a constrained realization of a single
large halo (see, for example, \pcite{springel01}), then the accretion
of smaller halos by this single large halo can be used to determine
the merger kernel $Q(M_1,M_2)$ for a single large $M_1$ as a function
of $M_2$ for $M_2\ll M_1$.  Such a numerical calculation may be
helpful in checking the numerical inversion we have carried out here,
and perhaps in shedding light on the asymptotic behaviour of
$Q(M_1,M_2)$ for $M_2 \ll M_1$. We plan to pursue such N-body
calculations in future work---initial investigations using simulations
with CDM power spectra suggest that our numerical results may better
match the N-body merger rates than the predictions of extended
Press-Schechter theory.

{\it What about fragmentation?}  In a realistic N-body simulation,
some of the mass in halos might be ejected (e.g., by an analog of the
``slingshot'' effect).  If so, then the coagulation equation should
include fragmentation terms.  We believe, however, that the fraction
of mass ejected will be small, and that the inexorable trend in
hierarchical structure formation is to larger masses; this is also
mathematically realized in the Press-Schechter and extended
Press-Schechter theories.  We therefore idealize the situation by
neglecting any such fragmentation and attempt to find a solution to
the pure coagulation problem first.

{\it What about ``smooth'' accretion?}  Some authors have noted that
merger trees assembled with the extended-Press-Schechter merger rates
do not produce Press-Schechter distributions, as they should, and have
thus attempted to account for this discrepancy by the supposition that
halos can gain mass by accreting diffuse matter from the smooth
intergalactic medium, as well as by mergers with other halos.  In our
formalism, what others would call ``smooth accretion'' is described by
mergers with very low-mass halos. In other words, every mass element
is contained in a halo of some size (as in standard Press-Schechter
theory), and smooth accretion is taken into account by mergers with
halos of extremely low mass.

{\it Is the numerically determined merger rate unique?} Given our
choice of regularization condition---namely that the function be as
smooth as possible and everywhere positive---we do indeed find a
unique solution. However, we could imagine other regularization
criteria, perhaps in addition to smoothness and positivity, which
would lead to different answers. Not all such criteria may allow a
solution to be found. For example, we find that requiring our
recovered merger rate to coincide with the extended Press-Schechter
merger rate for the case of equal mass mergers prevents us from
finding a merger rate which accurately solves the Smoluchowski
equation.

{\it Can the merger kernel be determined by looking at the properties
(e.g., size) of the physical halos?} In most applications of the
coagulation equation (e.g., in planetesimal theory), the merger kernel
is determined by the physical cross section and relative velocity of
the merging objects; i.e., by the ``micro''-physics.  In our case,
however, such a kinetic-theory description will not apply, as the
system of halos is ``cold'' and halos are more likely to merge with
those that form nearby, rather than a halo drawn at random from the
field.  In other words, the merger history is determined by the
clustering properties of the halos (as encapsulated by the power
spectrum) rather than by any intrinsic property, such as their size.

{\it What about other algorithms for generating merger trees?}
The fact that the ePS binary merger rates, when used to
construct merger trees, do not yield Press-Schechter
distributions is well known.  There have been algorithms
developed that mitigate the shortcomings of ePS by correcting
the ePS merger rates with, e.g., three-body mergers or
``smooth'' accretion (e.g. \pcite{sk99,cole2000}). There is no
unique algorithm for constructing a merger tree (\pcite{sk99} explore
many of the possibilities), but some are more successful than others
at reproducing the distribution of progenitors of dark-matter halos as
predicted by the extended Press-Schechter theory. It should be
noted, however, that the merger rates used in these merger trees
are not necessarily correct, even if the algorithm does
ultimately reproduce the correct mass function.  It should be
further noted that all of these algorithms rely on the extended
Press-Schechter merger rates for major mergers which, as we have
shown, are very likely incorrect.  As such, these algorithms do
not address the problem we consider here.  In fact, with our
self-consistent merger rates, merger trees can be constructed
without resorting to three-body mergers, smooth accretion, or
other kluges.

\section{Discussion}

We have described an approach to find physically reasonable estimates
of dark-matter-halo merger rates. We describe techniques for finding
numerical solutions to the Smoluchowski coagulation equation which we
use to describe the hierarchical formation of dark-matter halos. While
the extended Press-Schechter theory contains an intrinsic
inconsistency in its predictions for halo merger rates (namely that
the merger rate of halos of mass $M_1$ with those of mass $M_2$ is not
the same as that of halos of mass $M_2$ with halos of mass $M_1$) our
approach is guaranteed to always produce self-consistent merger rates.

We have presented symmetric, smooth solutions to the Smoluchowski
equation for the case of dark-matter--halo growth through gravitational
clustering. These solutions can now be checked against the results of
numerical simulations. The same techniques can also be applied to
non-power law power spectra (although in general the solution will be
somewhat more complex as the merger rate may depend explicitly on time
for such power spectra and may contain characteristic mass scales),
and also to halo-abundance evolution rates derived from fitting
functions designed to reproduce the halo mass functions found in
N-body simulations.

A number of questions still remain before we have reliable
astrophysical merger rates.  First and foremost among these is how to
insure that the numerically recovered solution is in fact {\it the}
solution that correctly describes mergers due to gravitational
amplification of an initially Gaussian distribution of
perturbations.  The fact that our numerical results are highly
insensitive to the details of our smoothing constraint leads us
to believe that our algorithm is indeed converging to a unique
physical result, but we cannot demonstrate this rigorously.
Perhaps future analytic work inspired by
extended-Press-Schechter--like considerations or N-body simulations
may provide insight into constraints, boundary conditions, and/or
asymptotic limits that will allow us to confidently zero in on the
correct solution.  Beyond that, the technique will then need to be
applied to obtain a merger kernel for a CDM power spectrum, rather
than the scale-invariant power spectra we have considered here.  And
finally, the merger kernel will ultimately have to be consistent with
a Sheth-Tormen or Jenkins et al. mass function, or whatever other mass
function is determined to be the ``correct'' one.  With these
self-consistent kernels, the well-known problems with ePS
formation-redshift distributions (i.e., that they are not
positive definite) will be corrected, and merger trees will be
easily constructed.  We plan to address these issues in future work.

The hierarchical formation of dark-matter halos is a fundamental
component of a large fraction of current studies in the fields of
cosmology and galaxy formation. A mathematically self-consistent and
quantitatively accurate knowledge of halo merger rates is therefore
extremely valuable. We believe that the techniques developed in this
work may provide a first step for developing such knowledge.

\section*{Acknowledgments}

We have benefited from discussions with many people about this work,
but acknowledge particularly useful early discussions with C.~Porciani
and S.~Matarrese.  MK acknowledges the hospitality of the Aspen
Center for Physics, where part of this work was completed.  This
work was supported at Caltech by NASA NAG5-11985 and DoE
DE-FG03-92-ER40701. AJB acknowledges a Royal Society University
Research Fellowship.

\appendix

\section{Alternative Formulation of the Coagulation Equation}
\label{ap:alternate}

Here we provide an alternative form for the coagulation equation that
demonstrates rigorously the cancellation between the divergences in
the creation and destruction terms.  This form of the equation is for
the time evolution for the fraction of the total mass ${\cal F}(<M)$
contained in halos of mass less than $M$.
This time evolution can be written,
\begin{eqnarray}
 {\d{\cal F}(<M)\over \d t} & &  = {1 \over 2\rho_0} \int_0^M\, \d
     M' \, M' \, \nonumber \\
     & &\times  \int_0^{M'} \d M_1 n(M_1) n(M'-M_1)
     Q(M_1,M'-M_1) \nonumber \\ 
     & & -  {1\over \rho_0}\int_0^M
     \d M'  M' n(M') \nonumber \\
     & & \times \int_0^\infty \d M_1 n(M_1)
     Q(M',M_1).
\label{eq:firstappendix}
\end{eqnarray}
The first step is to rewrite the first term on the right-hand
side as
\begin{eqnarray}
     {1\over 2}\int_0^M \, \d M' \int_0^\infty\, \d M_1
     \int_0^\infty \d M_2\, M' \delta(M_1+M_2-M') \nonumber \\
     \,\,\, \times n(M_1) n(M_2) Q(M_1,M_2).  
\end{eqnarray}
We then carry out the $M'$ integral which then changes $M'$ in
the integrand to $M_1+M_2$, and changes the upper limits to the
$M_1$ and $M_2$ integrals from $\infty$ to $M$ and $M-M_1$,
respectively; i.e., this expression is then
\begin{eqnarray}
     {1\over 2} \int_0^M \, \d M_1 \int_0^{M-M_1} \, \d M_2
     \, (M_1+M_2) \nonumber \\
     \times n(M_1) n(M_2) Q(M_1,M_2).
\end{eqnarray}
Since $M_1$ and $M_2$ enter symmetrically in this expression, it
can be replaced by
\begin{equation}
     \int_0^M \, \d M_1 \int_0^{M-M_1} \, \d M_2 \, M_1
      n(M_1) n(M_2) Q(M_1,M_2).
\end{equation}
We then make the replacements $M'\rightarrow M_1$ and
$M_1\rightarrow M_2$ in the dummy variables in the second term
in equation (\ref{eq:firstappendix}) and find that the first and
second terms are identical, except for in the limits for the second
integral.  Combining the two expressions, we find
\begin{eqnarray}
     \frac{d{\cal F}(<M)}{dt}& =& - \int_0^M\, \d M_1\, M_1
     n(M_1) \nonumber \\
     & & \times \int_{M-M_1}^\infty \, \d M_2\, n(M_2) Q(M_1,M_2).
\end{eqnarray}
It is then simple to see that if $n(M) \sim M^{-\gamma}$ with
$\gamma<2$ and $Q\sim$~constant as $M\rightarrow0$, then the divergence
in the integrand is integrable.

\section{Analytic Solutions to the Smoluchowski Equation}
\label{ap:an}

In this Appendix we provide an elementary approach to two known analytic
solutions to the Smoluchowski equation. These solutions are not new
(see \pcite{leyvraz} for a comprehensive review), but we include them
here in the hope that they may provide some insight into more general
solutions. We consider the models $q_{ij}=$constant, and
$q_{ij}=i+j$.  The first leads to an exponential
mass function for large times while  
the second gives the Press-Schechter mass function for
an $n=0$ power spectrum.

The Smoluchowski equation is
\begin{eqnarray}
\dot{n}_1 & = & - \sum_{i=1}^\infty q_{i,1} n_i n_1 , \nonumber \\
\dot{n}_j & = & {1 \over 2} \sum_{i=1}^{j-1} q_{i,j-i} n_i n_{j-i} -
\sum_{i=1}^\infty q_{ij} n_i n_j ,
\label{eq:apsmol}
\end{eqnarray}
where $q_{ij}$ corresponds to the discretized coagulation kernel.

We will consider solutions which satisfy the initial conditions
\begin{equation}
n_1=1, \quad n_r=0 \quad r \neq 1.
\end{equation}
It will also be useful to define
\begin{equation}
M_p(t) = \sum_{j=1}^\infty j^p n_j(t);
\end{equation}
i.e., the various moments of the number distribution, as well as the following
generating function,
\begin{equation}
F(z,t)=\sum_{j=1}^\infty n_j(t)\exp(-jz).
\end{equation}

We now consider each case individually.
\smallskip

{\it Constant kernel}: $q_{ij}=$ const. Inserting the
generating function into equation~(\ref{eq:apsmol}) gives
\begin{equation}
{\partial F \over \partial t} = {F^2 \over 2}  - F M_0(t).
\label{eq:apdc}
\end{equation}
The initial conditions imply that 
$F(z,0)=\exp(-z)$, and setting
$z=0$ in equation~(\ref{eq:apdc}), 
gives $M_0(t)=2/(t+2)$. The generating function can then be solved for:
\begin{equation}
F(z,t)={4 \over (t+2)^2\exp(z)-t(t+2)}.
\end{equation}
Expanding this gives
\begin{equation}
n_k(t) = {4 \over (t+2)^2} \Big( {t\over (t+2)}\Big) ^{k-1}. 
\label{eq:apsol}
\end{equation}
The asymptotic behaviour of equation~(\ref{eq:apsol}) is of the form
\begin{eqnarray}
     n_k &\sim & t^{-2}g(k/t),\qquad  t\rightarrow \infty,\nonumber \\ 
     k&=& {\mathcal O}(t),\qquad g(\eta)=4\exp(-2\eta).
\end{eqnarray}

\smallskip

{\it Sum kernel}: $q_{ij} = i + j$. Plugging in for the generating function 
here gives 
\begin{equation}
{\partial F \over \partial t} =  {1\over 2}{\partial F\over \partial z}M_0(t)- {F \over 2}  {\partial F\over \partial z} - {F\over 2}M_1(t).
\label{eq:apdc2}
\end{equation}
Setting  $z=0$ in equation~(\ref{eq:apdc2})
gives $M_0(t)=\exp(-t/2)$, and assuming $M_1(t) \equiv 1$, we have, via the 
technique of characteristics (\pcite{leyvraz}),
\begin{equation}
{\rm e}^{-z} = F \exp\left(1-{\rm e}^{-t/2}+{t\over 2}\right)
\exp\left[ \left({\rm e}^{-t/2}-1 \right)F \right].
\end{equation}
The solution is then given using Lagrange's expansion \cite{handbook},
\begin{equation}
n_k(t) = {k^{k-1}\over k!}{\rm e}^{-t/2}\exp[-k(1-{\rm e}^{-t/2})](1-{\rm
e}^{-t/2})^{k-1}.
\label{eq:apnogel}
\end{equation}

The large-time behaviour can easily be extracted from
equation~(\ref{eq:apnogel}) as
\begin{equation}
n_k \sim {k^{k-1}\over k!} {\rm e}^{-k} {\rm e}^{-t/2},\qquad
t\rightarrow\infty,\qquad k={\mathcal O}(1),
\end{equation}
and, from Stirling's formula, as
\begin{eqnarray}
n_k &\sim & {\rm e}^{-2t}g(k/{\rm e}^t),\qquad
t\rightarrow\infty,\nonumber \\
k&=&{\mathcal O}({\rm e}^t),\qquad g(\eta) = {{\rm e}^{-\eta/2}\over
\sqrt{2\pi\eta^3}}.
\end{eqnarray}


\begin{thebibliography}{}

\bibitem[Abramowitz \& Stegun<1974>]{handbook}Abramowitz~M. and Stegun~I.~A. eds., 1974, ``Handbook of Mathematical Functions'', Dover Publications

\bibitem[Aldous<1999>]{Ald99} Aldous D. J., 1999, Bernoulli, 5, 3

\bibitem[Allen \& Bastien<1995>]{AllBas95}  Allen E. J., Bastien
     P., 1995, ApJ, 452, 652

\bibitem[Benson et al.<2002>]{Benetal02} Benson~A.~J., Frenk~C.~S., Lacey~C.~G., Baugh~C.~M., Cole~S., 2002, MNRAS, 333, 177

\bibitem[Bond et al.<1991>]{Bonetal91} Bond~J.~R., Cole~S., Efstathiou~G., Kaiser~N., 1991, ApJ, 379, 440

\bibitem[Bond \& Myers<1996>]{bond96}Bond~J.~R., Myers~S.~T., 1996, ApJS, 103, 63

\bibitem[Bower<1991>]{Bow91} Bower R. G., 1991, MNRAS, 248, 332.

\bibitem[Brent <1973>]{Bre73} Brent~R.~P., 1973, ``Algorithms for
     Minimization without Derivatives'', Prentice-Hall,
     Englewood Cliffs, New Jersey

\bibitem[Bromm, Coppi \& Larson<1999>]{Bro99} Bromm V., Coppi P. S.,
     Larson R. B., 1999, ApJ, 527, L5

\bibitem[Bullock, Kravtsov \& Weinberg<2000>]{BulKraWei00}
     Bullock J. Kravtsov A. Weinberg D. H., 2000, ApJ, 539, 517

\bibitem[Cole et al.<2000>]{cole2000}Cole~S., Lacey~C.~G., Baugh~C.~M., Frenk~C.~S., 2000, MNRAS, 319, 168

\bibitem[Gabici \& Blasi<2003>]{GabBla03} Gabici S., Blasi
     P., 2003, ApJ, 583, 695

\bibitem[Gottlober, Klypin \& Kravtsov<1991>]{GotKlyKra99}
     Gottlober S., Klypin A.,Kravtsov A., 1999, ApSS, 269, 345

\bibitem[Haehnelt<1994>]{Hae94} Haehnelt M. G., 1994, MNRAS,
     269, 1999

\bibitem[Jenkins et al.<2001>]{Jenetal01} Jenkins A. et al.,
     2001, MNRAS, 321, 372

\bibitem[Kamionkowski \& Liddle<2000>]{KamLid00} Kamionkowski M., Liddle
     A. R., 2000, PhRvL, 84, 4525

\bibitem[Kolatt et al.<1999>]{Koletal99} Kolatt~T.~S. et al.,
     1999, ApJ, L523, 109

\bibitem[Lacey \& Cole<1993>]{LacCol93} Lacey C. G., Cole S.,
     1993, MNRAS, 262, 627

\bibitem[Lacey \& Cole<1994>]{LacCol94} Lacey C. G., Cole S.,
     1994, MNRAS, 271, 676.

\bibitem[Lee<2000>]{Lee00} Lee M. H., 2000, Icarus, 143, 74

\bibitem[Leyvraz<2003>]{leyvraz} Leyvraz F., 2003, Phys. Rept., 383, 95

\bibitem[Malyshkin \& Goodman<2001>]{MalGoo01} Malyshkin
     L. Goodman J., 2001, Icarus, 150, 314

\bibitem[Menou, Haiman \& Narayanan<2001>]{MenHaiNar01} Menou
     K., Haiman Z., Narayanan V. K., 2001, ApJ, 558, 535

\bibitem[Milosavljevic \& Merritt<2001>]{MilMer01} Milosavljevic M.,
     Merritt D., 2001, ApJ, 563, 34

\bibitem[Monaco et al.<2002>]{Monetal02} Monaco~P., Theuns~T., Taffoni~G., Governato~F., Quinn~T., Stadel~J., 2002,
     ApJ, 564, 8

\bibitem[Muralidar \& Ramkrishna<1986>]{MurRam86} Muralidar R., Ramkrishna
     D., 1986, J. Colloid and Interface Sci., 112, 348

\bibitem[Percival et al.<2003>]{Peretal03} Percival W. J., Scott
     D., Peacock J. A., Dunlop, J. S., 2003, MNRAS, L338, 31

\bibitem[Press \& Schechter<1974>]{PreSch74} Press W.,
     Schechter P., 1974, ApJ, 187, 425

\bibitem[Santos, Bromm \& Kamionkowski<2002>]{SanBroKam02} Santos M. R.,
     Bromm V., Kamionkowski M., 2002, MNRAS, 336, 1082

\bibitem[Scannapieco, Schneider \& Ferrara<2003>]{ScaSchFer03}
     Scannapieco E., Schneider F., Ferrara A., ApJ, 589, 35

\bibitem[Sheth<1995>]{She95} Sheth R., 1995, MNRAS, 274, 213

\bibitem[Sheth \& Pitman<1997>]{ShePit97} Sheth R. K., Pitman
      J., MNRAS, 1997, 289, 66

\bibitem[Sheth \& Tormen<1999>]{SheTor99} Sheth R. K., Tormen
     G., 1999, MNRAS, 308, 119

\bibitem[Silk \& Takahashi<1979>]{SilTak79} Silk J.,
     Takahashi T., 1979, ApJ, 229, 242

\bibitem[Silk \& White<1978>]{SilWhi78} Silk J., White S. D. M.,
     1978, ApJ, 223, L59

\bibitem[Smoluchowski<1916>]{Smoluchowski} Smoluchowski M., 1916, Physik. Zeit., 17, 557

\bibitem[Somerville<2002>]{Som02} Somerville R. S., 2002, ApJ, 572, 23

\bibitem[Somerville \& Kolatt<1999>]{sk99}Somerville~R.~S., Kolatt~T.~S., 1999, MNRAS, 305, 1

\bibitem[Springel et al.<2001>]{springel01}Springel~V., White~S.~D.~M., Tormen~G., Kauffman~G., 2001, MNRAS, 328, 726

\bibitem[Stiff, Widrow \& Frieman<2001>]{StiWidFri01} Stiff D., Widrow L. M.,
     Frieman J., 2001, PhRvD, 64, 3516

\bibitem[Verde et al.<2001>]{Veretal01} Verde L., Kamionkowski
    M., Mohr J., Benson A. J., 2001, MNRAS, 321, L17

\bibitem[Verde, Haiman \& Spergel<2002>]{VerHaiSpe02} Verde L.,
     Haiman Z., Spergel D. N., 2002, ApJ, 581, 5

\bibitem[Vivitska et al.<2002>]{Vivetal02} Vitvitska~M., Klypin~A.~A., Kravtsov~A.~V., Wechsler~R.~H., Primack~J.~R., Bullock~J.~S., 2002, ApJ, 581, 799

\bibitem[Volonteri, Haardt \& Madau<2002>]{VolHaaMad02}
     Volonteri M., Haardt F., Madau P., 2002, Ap\&SS, 281, 501

\bibitem[Wechsler et al.<2002>]{Wecetal02} Wechsler~R.~H.,
     Bullock~J.~S., Primack~J.~R., Kravtsov~A.~V., Dekel~A., 2002,
     ApJ, 568, 52

\bibitem[Wetherill<1990>]{Wet90} Wetherill G. W., 1990, Icarus, 88,
     336

\bibitem[White<2002>]{Whi2002} White M., 2002, ApJS, 143, 241

\bibitem[Wyithe \& Loeb<2003>]{WyiLoe03} Wyithe S., Loeb A.,
     2003, ApJ, 595, 614

\end{thebibliography}
\end{document}